\documentclass[conference]{IEEEtran}

\usepackage{cite}

\ifCLASSINFOpdf
  \usepackage{graphicx}
\else
\fi

\usepackage{url}

\newcommand{\sysname}{{\textit{Stitch}}}
\newcommand{\spotlight}{\textit{Spotlight}}
\newcommand{\pctl}{\textit{P4Control}}

\usepackage{subcaption}
\usepackage[table]{xcolor}
\usepackage{comment}

\begin{document}

\title{Effective Pivot Attack Detection via System and Network Information}

\author{\IEEEauthorblockN{Ava Powelson}
	\IEEEauthorblockA{Dalhousie University\\
		apowelson@dal.ca}
	\and
	\IEEEauthorblockN{Carson Kuzniar}
	\IEEEauthorblockA{Dalhousie University\\
		carson.kuzniar@dal.ca}
	\and
	\IEEEauthorblockN{Hyojoon Kim}
	\IEEEauthorblockA{University of Virginia\\
		joonkim@virginia.edu}
	\and
	\IEEEauthorblockN{Israat Haque}
	\IEEEauthorblockA{Dalhousie University\\
		israat@dal.ca}}	

\IEEEoverridecommandlockouts
\makeatletter\def\@IEEEpubidpullup{6.5\baselineskip}\makeatother

\maketitle

\begin{abstract}
Perimeter-based security appliances, such as firewalls or Intrusion Detection Systems, are ineffective against modern attacks that use pivoting, wherein attackers “pivot” traffic through compromised hosts to gain access to additional targets that would otherwise be inaccessible. Due to the legitimate appearance of the relayed traffic, pivoting is extremely difficult to detect. Although the consequences of these attacks are known to be severe, existing defenses suffer from drawbacks such as high processing delays, low accuracy, or reliance on network-wide participation, making them inconvenient or even ineffective. This work presents \sysname{}, a host-based system that uses the programmable kernel to detect pivoting in real time. By observing host-traversing flows, \sysname{} uses process tracing to effectively combine system and network-level information, connecting incoming and outgoing communications and identifying pivoting characteristics between them. Showing 31\% gains in accuracy over state-of-the-art pivoting defenses and a maximum false-positive rate of 0.006\% over two separate real-world deployments, \sysname{} covers the gap in current pivot detection solutions by providing accurate, lightweight, and independent coverage for vulnerable hosts in a network.
\end{abstract}

\IEEEpeerreviewmaketitle

\section{Introduction}
\label{sec:introduction}
In today’s cyber-security landscape, critical infrastructure is often targeted by Advanced Persistent Threat (APT) attacks \cite{aptnotes_2025, fortinet_apts}. Causing extensive loss of capital, infrastructure, and services \cite{aptnotes_2025, fortinet_apts, nl_health, us_pipeline, trellix}, these attacks demonstrate a concrete threat to both individual and organizational privacy \cite{trellix}. APTs have tangible and global negative effects on user experience, privacy, and providers' services. As such, solutions to detect, mitigate, and prevent these attacks are critical.

\begin{figure}
    \centering
    \includegraphics[width=\linewidth]{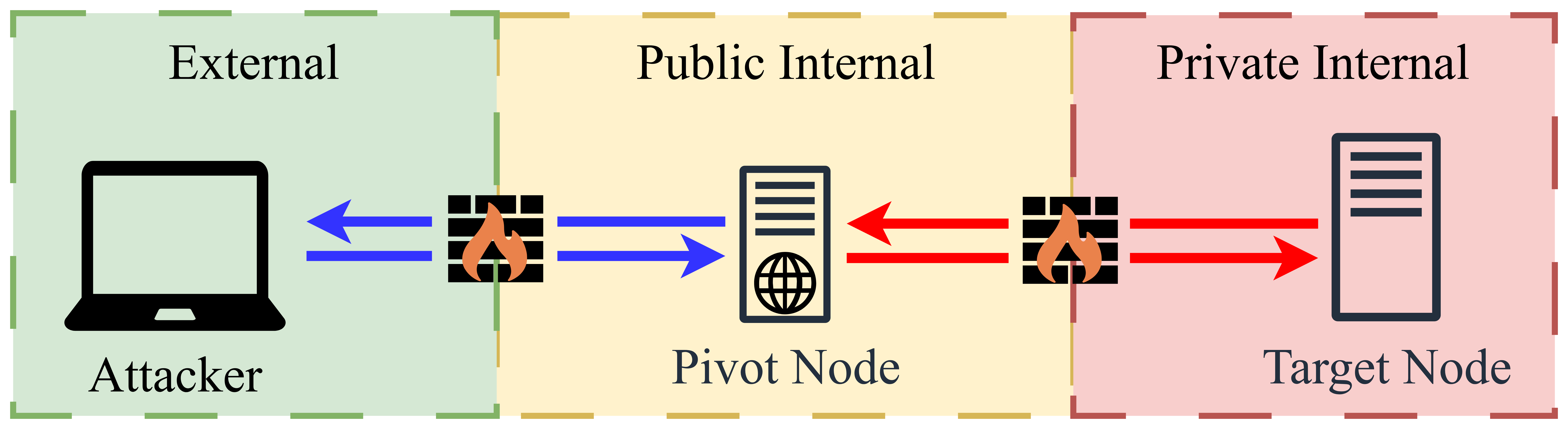}
    \caption{An example of a pivot attack. The attacker "pivots" their traffic through a public internal node and exploits the trusted internal traffic to access the private internal node.}
    \label{fig:pivot_example}
\end{figure}

One common facet of APT attacks' presence in a network is \emph{lateral movement} \cite{apt_survey}. This type of movement defines communications between internal network devices, as opposed to \emph{vertical movement} which defines communications between internal and external devices. Traditionally, security solutions such as Firewalls and Intrusion Detection Systems (IDS) focus on detecting and preventing unauthorized vertical movement while trusting internal communications on the assumption that internal hosts will emit only legitimate traffic \cite{mitre_lat}. As a result, a technique called \emph{pivoting} (Fig.~\ref{fig:pivot_example}) has emerged to take advantage of this relatively unmonitored lateral traffic. Pivoting is a stealthy movement technique in which attackers utilize the more trusted communications of internal network devices by relaying or ``pivoting'' their traffic through one or more such devices to communicate with their target (or targets) without raising suspicion \cite{apt_survey, mitre_lat}.

Preventing pivoting is an essential step in defending against APTs. If attackers' lateral movements are detected and restricted, they will be forced to utilize vertical movement instead and contend with existing perimeter defenses. However, detecting pivoting poses significant challenges due to the legitimate appearance of its traffic and the diversity and volume of benign activity in today's networks. Infrastructure as a Service (IaaS) and Platform as a Service (PaaS) paradigms further complicate the issue by introducing dynamic network scenarios where certain elements, such as applications or resource allocation, fall outside of administrator control.

Existing work has proposed several different solutions to the pivoting problem \cite{hopper, apivads, apruzzese, spotlight, p4control, pivotwall}. For example, some solutions \cite{hopper, apruzzese, prographer} rely on a central server to collect and process traffic for \textit{offline pivot detection}, but this strategy often suffers from delayed responses and large data-storage requirements when processing high traffic volumes. Other solutions implement end-to-end Distributed Information Flow Control (DIFC) systems, which track and control the flow of information in a given network to prevent unauthorized lateral movement \cite{pivotwall, p4control}. However, such systems rely on the \textit{cooperation of all nodes in the network}, including hosts, servers, switches, etc., to ensure proper rule enforcement, making them incompatible with dynamic network topologies such as bring-your-own-device (BYOD), IaaS, or PaaS environments. Finally, some solutions analyze flow characteristics, such as their size and timing, to perform lightweight real-time detection \cite{spotlight, apivads}. However, relying on such characteristics alone does not provide enough context, leading to the tracking of many unrelated flows and resulting in a \textit{high false-positive rate} (FPR).

A pivot detection system must operate in real-time and scale in the presence of a high traffic volume without relying on the cooperation of existing network elements, all while minimizing the FPR. This combination is missing in existing work (Section \ref{sec:related}). \sysname{} closes the identified gaps by combining flow characteristics (e.g., size and time) analysis and system-level context (e.g., flow association via Information Flow Control (IFC)) in a principled, lightweight approach. Our key insight is that host-based pivot detection must correlate network flows with serving system processes and, as such, consider full system context along with traffic characteristics to effectively identify potential pivoting. For instance, tracking an SSH tunnel's flow characteristics in isolation is insufficient without considering the associative context linking its incoming and outgoing flows through shared process activity.

Realizing this insight introduces several technical challenges (Section \ref{subsec:chals}) that shape \sysname{}'s design. First, legitimate symmetric traffic patterns are easily mistaken for pivoting, so any system relying on flow symmetry alone risks burying real alerts under false ones. \sysname{} therefore learns which endpoint paths are frequently traversed from a host, inferring which flows are likely to be administrator-managed and can be safely ignored. Second, a host-resident monitor also has only a narrow budget of compute and storage resources to spend without degrading the system it protects, ruling out heavyweight instrumentation or unbounded state. \sysname{} therefore couples targeted kernel-level monitoring with an eviction strategy that exploits the temporal stability of pivoting symmetry to keep its footprint small without losing detection accuracy. Finally, the convoluted, multi-threaded data paths of modern OS processes can threaten IFC schemes with label explosion. \sysname{} solves this issue by attributing child thread activity to parent processes, preserving the context needed for pivot detection while avoiding the storage and complexity costs of per-thread tracking. No prior system addresses all three constraints simultaneously, making \sysname{} a uniquely accurate, real-time, and lightweight system.

To realize these solutions, \sysname{} leverages eBPF (extended Berkeley Packet Filter), a framework for safely and dynamically extending the Linux operating-system (OS) kernel \cite{ebpf_io}. \sysname{} deploys kernel-level programs at carefully chosen system-call monitoring points, i.e., critical stages in a process's lifespan such as \textit{execve} and \textit{clone}, intersections of process and flow activity such as \textit{connect} and \textit{accept}, and calls as close possible to the network via eXpress Data Path (XDP). This design directly addresses the resource-constraint challenge by extracting rich system and network data without excessive instrumentation or user/kernel context-switching costs. 

The collected data further drives two key mechanisms: an IFC scheme performed at the process level, which attributes individual threads' actions to their parent to sidestep the convoluted-data-path challenge without any loss of detection-relevant context, and a false-positive reduction strategy built on the endpoints a protected host frequently visits, allowing \sysname{} to infer well-defined, administrator-managed flows (e.g., authentication management or maintenance traffic) without flagging every instance of symmetric traffic. By unifying these solutions, \sysname{} offers a single system that performs lightweight, real-time, and accurate pivot detection entirely within the kernel.

We implement a prototype of \sysname{} and evaluate it on traffic modeled after the Canadian Institute for Cybersecurity Intrusion Detection System 2017 (CIC-IDS2017) dataset and augmented with several representative pivoting attacks (Nmap, SSH, Socat, and Chisel) selected for their prevalence in prior work \cite{spotlight, p4control, apivads} and in real-world security incidents \cite{afr_fin, financial_atks, pysa}. In this testbed setting, \sysname{} identifies all representative attacks while incurring only 0.1\% additional CPU overhead and a memory footprint under 350MB. We further validate \sysname{} with two live production deployments in the campus network of a large university (24,000+ students, faculty, and staff) for over 80 days. For these deployments, \sysname{} was placed first in a shared, multi-purpose faculty server and second in a public-facing university web server. \sysname{} sustained a maximum false positive rate of just 0.006\% over the combined three months of these deployments. The high accuracy of \sysname{} and its low impact on host processing resources establish its feasibility as a real-world pivot detection tool. Our contributions are summarized as follows:

\begin{itemize}

\item \sysname{} is the first real-time host-based pivot detection system combining process tracing with network flow analysis, enabling efficient and accurate detection at scale without relying on full network cooperation.

\item We extensively evaluate \sysname{} across three distinct setups (a controlled testbed and two live production networks) demonstrating that it is highly accurate, lightweight, and portable across different kernel versions.

\item We compare \sysname{} with state-of-the-art pivot detection methods and demonstrate an average 31.49\% improvement in detection accuracy while incurring an average FPR of only 0.18\% in the testbed environment.
  
\item We will publicly release \sysname{} source code upon acceptance of this paper to support reproducibility and enable future extensions.

\end{itemize}

\section{Background and Motivation}
\label{sec:background}

This section first presents a detailed pivot attack example, followed by motivating insights into inferring flow causality and the role of eBPF, and finally describes the threat model.

\textbf{Pivoting.} Pivoting is carried out by compromising (through malware or sensitive information leakage) one or more hosts in the target network and using them to ``pivot'' traffic to other nodes. By pivoting traffic through compromised nodes, the attacker gains valid IP addresses and, from them, access to target hosts that would otherwise be inaccessible due to traditional safeguards. 

Fig.~\ref{fig:pivot_example} shows an example pivot scenario, which could progress as follows: first, a remote attacker gains access to the pivot node by using credentials gained through social engineering methods or by compromising an insecure web interface, such as in CVE-2025-48703 \cite{CVE-2025-48703}. With access established to this intermediary node, the attacker could then gain additional network information through unprivileged reconnaissance tools, such as Nmap \cite{nmap}. Then, even with unprivileged access, there are a number of common tools that allow users to create outgoing connections from a pivot node, including SoCat \cite{socat}, Chisel \cite{chisel}, and SSH \cite{ssh}. After creating a connection between the pivot node and a target node using one of these tools, the pivot node can be used for both propagation and extraction of data, meaning that an attacker could spread malware to the internal target node or exfiltrate private data to their remote machine. Additionally, the attacker could continue to move laterally into the network if further links exist between the target and other internal nodes. Such unauthorized access can result in sensitive data disclosure, have negative financial repercussions, and cause disruptions for both clients and administrators \cite{mitre_lat}. 

\subsection{Why must network and system data meet to infer causality?}\label{subsec:causality}

Existing solutions implement one of two design principles: system-level \textit{process tracing} \cite{p4control, pivotwall} or network-level \textit{flow characteristic analysis} \cite{apivads, spotlight}. We discuss the issues inherent with these two strategies and position \sysname{} to integrate the strengths of both schemes, leveraging their individual benefits in a novel pivot detection strategy.

\begin{figure}
    \centering
    \includegraphics[width=\linewidth]{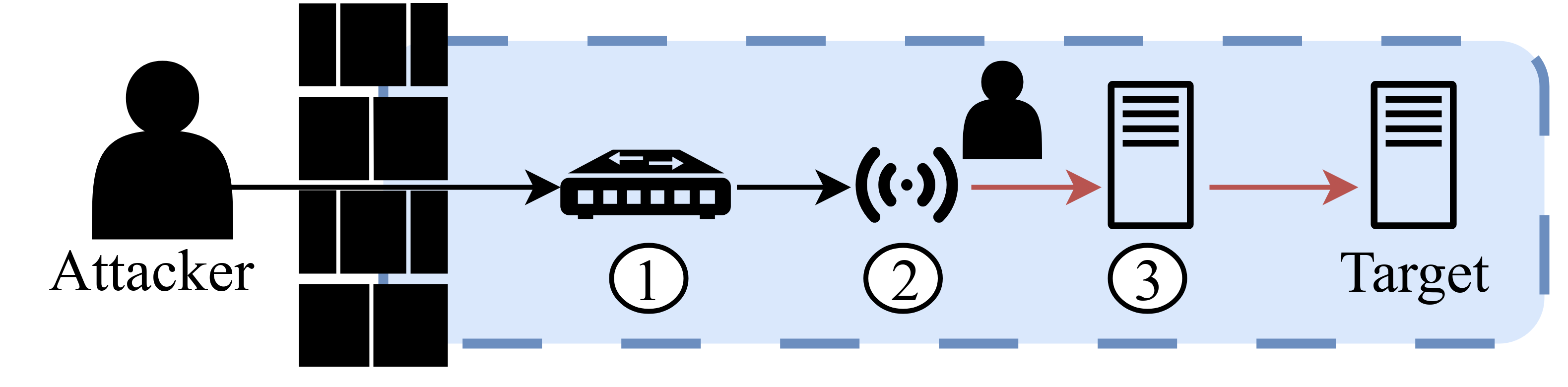}
    \caption{A dynamic network scenario in which the traffic route has encountered an off-policy device in a DIFC controlled network, allowing for unauthorized lateral movement.}
    \label{fig:dynamic_net}
\end{figure}

\textbf{Process tracing.} Process tracing attaches labels to target processes in order to track associated user activities and infer causality. In network-wide Distributed Information Flow Control (DIFC) systems, these labels are attached to the outgoing packets of a connection to infer associated activities across different machines. However, process tracing in DIFC can be unreliable when encountering non-participating nodes, as shown in Fig.~\ref{fig:dynamic_net}. In this example, a pivoting flow in an enterprise network has gone undetected due to the presence of a non-policy-enforcing (off-policy), user-supplied device (\raisebox{.5pt}{\textcircled{\raisebox{-.9pt} {2}}}) through which labeled traffic has been routed. In this case, the device could be an access point, such as an off-the-shelf WiFi extender, that does not comply with enterprise policies. Off-policy or BYOD devices may drop or ignore important DIFC labels and forward traffic that would otherwise be stopped: since the DIFC policy enforcement on \raisebox{.5pt}{\textcircled{\raisebox{-.9pt} {3}}} relies on labels propagated from \raisebox{.5pt}{\textcircled{\raisebox{-.9pt} {1}}}, it is rendered ineffective if those labels are lost or corrupted en route. As \sysname{} does not rely on network-wide cooperation, deploying it on \raisebox{.5pt}{\textcircled{\raisebox{-.9pt} {3}}} would ensure detection of pivoting traffic crossing the node, even if the preceding path is modified. \sysname{} is designed to be deployed on individual nodes that could be stepping stones for pivot attacks (i.e., pivot nodes) and can be implemented in a complementary fashion with other security solutions. 

In contrast, host-based works that use process tracing tend to focus on system auditing for after-the-fact intrusion analysis \cite{protracer,eaudit, camflow, provbpf}. While valuable, these approaches have large storage requirements, long wait times before attacks are identified, and high implementation overheads, making them unsuitable for real-time pivot detection.

\textbf{Flow characteristics.} In this case, causality between flows is inferred by inspecting their \textit{timing} and \textit{size} characteristics, i.e., if an incoming flow begins within a certain time frame before an outgoing one and their sizes are similar, it is considered as a pivoting relationship \cite{spotlight, apivads}. This scheme relies on the assumption that commands or information propagated across a host machine will result in nearly identical content in the incoming and outgoing flows, leading to nearly identical sizes. It also assumes that an occurrence of two such flows within a small time window is rare in non-tunneling traffic, and therefore establishes causality based on that timing. This methodology can be applied in-network via programmable switches \cite{spotlight} or as a host-based approach \cite{apivads}.

However, when pivoting inference relies solely on flow-characteristics, unrelated flows are frequently tracked due to timing or a high traffic volume where independent flows are detected within the same time window. This can generate a large number of false positives and negatively affects the practical deployability of such tools \cite{apivads, spotlight}. 

\subsection{How can eBPF help?} \label{subsec:ebpf}

eBPF is a rapidly growing technology that allows developers to apply customized, dynamic, and safe extensions to the Linux operating system (OS) kernel \cite{ebpf_io}. Its event-based programs are attached to kernel level \emph{hook points}, i.e., pre-defined triggers for program execution. Because they run in a privileged context, eBPF programs are statically inspected by a \textit{verifier} before being loaded into the kernel. The verifier prevents the loading of any potentially damaging eBPF programs and ensures that executed programs are kernel safe. eBPF programs are ephemeral and use eBPF maps to store state and communicate between programs. These maps are data structures that can be shared between any number of eBPF programs. They also provide efficient methods of communication between kernel and userspace programs \cite{bpf_maps}. 

By hooking into key locations in the kernel of a potential pivot node, \sysname{} uses eBPF to perform efficient pivot detection for any individual network host. Process tracing can be performed via process lifespan events such as \textit{execve}, \textit{clone}, or \textit{fork}, while the necessary network information can be gathered and analysed at packet ingress and egress. Using eBPF maps, these monitoring components can easily share data and coordinate between the network and system domains. In fact, the entire pivot detection pipeline can be implemented \textit{inside the kernel}; no interaction with userspace is required until pivoting is detected, wherein \sysname{} will emit an alert to the user. Thus, \sysname{} avoids costly context switching between user and kernel space to respond rapidly to pivoting while maintaining a small footprint on system resources.

\subsection{Threat Model} \label{subsec:threat_model}
We consider a single, network-connected pivot node and make three important assumptions. First, it is assumed that the attacker cannot directly connect to their target node (e.g., due to traditional firewall) and must use the accessible pivot node as a relay. This type of setup, an example of which is shown in Fig.~\ref{fig:pivot_example}, occurs frequently in real-world deployments such as web servers or cloud providers, where only certain resources in a network are publicly exposed: an exterior firewall forms an initial perimeter with relaxed access rules for public resources, while an internal one creates a more strict perimeter around private network resources. Second, similar to other pivot detection works with host based elements \cite{pivotwall, apivads, p4control}, it is assumed that the attacker has compromised the pivot node but does not have root access. Finally, we assume that the attacker is relaying traffic across the pivot node without delay, as in \cite{spotlight, hopper}. For example, the attacker in Fig.~\ref{fig:pivot_example} would establish the secondary connection to the target node immediately upon access to the pivot node. \sysname{} does not require any additional assumptions about network topology or the type of node on which it is deployed, making it widely compatible with different topologies and systems.

\section{Methodology and Design}
\label{sec:system}

We first present design challenges and principles of \sysname{}, followed by its workflow, detailed architecture, and operation.

\subsection{Challenges and Principles}\label{subsec:chals}

\textbf{False Positive Reduction.} One challenge for pivot detection is that legitimate network activity, such as port-forwarding and other symmetric traffic patterns (i.e., similar on incoming and outgoing paths), frequently resembles pivot attacks closely enough to be falsely flagged by pivot detection tools \cite{apivads, hopper, spotlight, apruzzese}. Excessive false positive alerts that require manual intervention can cause administrator fatigue and reduce the overall effectiveness of detection tools by burying legitimate alerts under superfluous ones. 

To reduce false-positive alerts that arise from such activity and reduce manual intervention, we observe (confirmed in Section \ref{sec:evaluation}) that many of these alerts come from well-defined, administrator managed flows (e.g., authentication management or maintenance traffic). Thus, we adopt a frequency analysis strategy, tracking which endpoints are commonly accessed from a given host to build an understanding of the well-defined paths crossing that host (Section \ref{subsec:ep_conn_analysis}). Such paths are likely to be monitored by administrators and other management tools as a part of a defense-in-depth strategy, allowing \sysname{} to recognize them as non-anomalous and forego excessive alerting on such common paths. This approach allows \sysname{} to effectively reduce false positives with a high true-positive detection accuracy and a lightweight footprint on its host OS.

\textbf{Limited access to information and resources.} 
In order to yield computing and storage resources to the core functions of its host system, \sysname{} must employ a lightweight detection scheme and judiciously store relevant data. Enabled by eBPF, the carefully chosen kernel-level monitoring points provide rich system data and allow \sysname{} to gather the information necessary to detect pivoting \textit{without} excessive instrumentation (Section \ref{subsec:info_flow_tracing}). That is, we choose hook points that fall at critical stages in process lifespans (such as \textit{execve} or \textit{clone}), at intersections of process and flow activity (such as \textit{connect} or \textit{accept}), and those that fall as close to the network as possible (such as the \textit{eXpress Data Path (XDP)}). These chosen monitoring points enable lightweight, accurate, and efficient pivot detection. Further, \sysname{} deploys its detection pipeline entirely within the kernel, allowing it to capitalize on this rich information while avoiding costly context switching between user- and kernel-space.

Additionally, \sysname{} implements rigorous eviction methods for all stored data (Section \ref{subsec:storage_management}). By recognizing that the symmetry of a pivoting flow pair remains stable throughout its lifetime and that pivoting can be identified within a very small time frame, we can implement smaller data stores with higher turnovers without compromising detection rates. Specifically, consistent flow symmetry means that even if contextual information about a pivoting flow is preemptively evicted, subsequent packets will in turn evict older flows' data and restore the current flows' context, providing additional opportunities for pivot detection throughout the lifetime of the connection. This strategy allows \sysname{} to maintain a robust system and network context with low storage requirements, and is confirmed by evaluation in Section~\ref{subsec:scalability}.

\textbf{Convoluted data-paths.} Modern operating systems often coordinate execution of hundreds of multi-threaded processes at once, with each thread handling different portions of the parents' responsibilities. This means that the flow of information between processes or threads is rarely straightforward, and tracking information flow with labels can easily lead to label explosion and excessive meta-data storage requirements. When process tracing, \sysname{} mitigates this issue by relying on the dependency of a child thread on its parent process. That is, the actions of child threads do not need to be tracked individually, but can instead be attributed to their parent process without a loss of information. This approach reduces active monitoring requirements and in turn reduces the amount of storage required for IFC tasks. Process labeling at the parent level is sufficient for pivoting context creation and significantly reduces resource requirements without losing relevant contextual information. 

\subsection{System Overview}
\begin{figure*}
    \centering
    \includegraphics[width=1\linewidth]{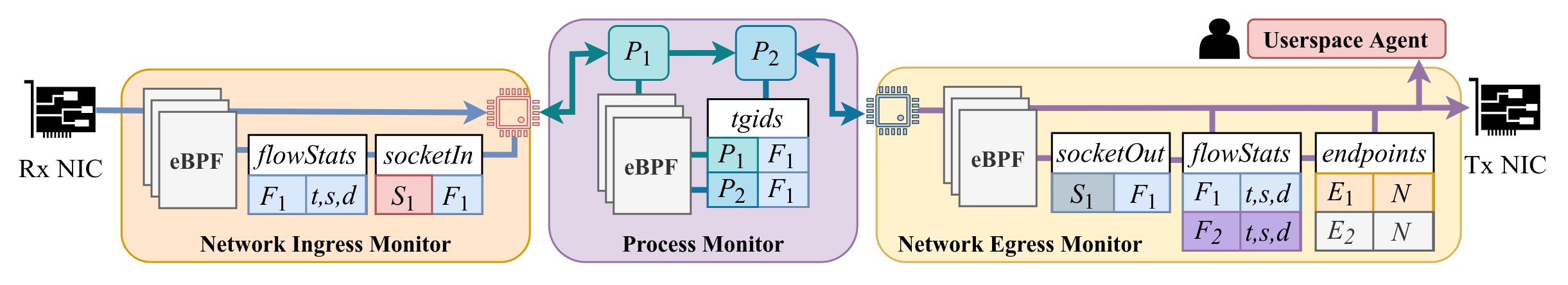}
    \caption{Complete \sysname{} architecture. The system uses eBPF programs and maps to propagate incoming-flow labels through the machine. Tracking starts with the incoming flow and its receiving socket, then on to the receiving process and its transmitting sockets, and finally to any outgoing flows that occur as a result of that same process or its children.}
    \label{fig:full_sys}
\end{figure*}

\sysname{} first uses process tracing to establish causality between incoming and outgoing flows, then analyzes their sizes, timing, and destination to determine if they exhibit pivoting characteristics. Applied as a series of filters, these steps efficiently identify pivot attacks while discarding consideration of legitimate tunneling traffic. The system consists of four components: the \textit{Userspace Agent}, the \textit{Network Ingress Monitor}, the \textit{Process Monitor}, and the \textit{Network Egress Monitor}, which are illustrated in Figure~\ref{fig:full_sys}. 

\textbf{Components.} Upon initialization, the \textit{Userspace Agent} loads the detection system code (eBPF), maps it into the kernel, and then processes any detection results received from the kernel. The \textit{Network Ingress Monitor} monitors network ingress points on the host machine. Specifically, it tracks incoming network packets and maintains up-to-date records of every ongoing flow. It also labels the associated receiving sockets with the corresponding incoming-flow ID for subsequent identification by the \textit{Process Monitor}.

The Process Monitor is responsible for tracking the life cycle and activity of any process that receives network traffic. It begins monitoring a process once that process has accessed an incoming connection on a labeled socket. After the socket's label is assigned to the process, it is then monitored for its remaining lifespan, with any children inheriting its labels. 

Finally, the \textit{Network Egress Monitor} is responsible for monitoring network egress points and performing pivot detection. Similar to the Ingress Monitor, it tracks outgoing packets and updates the corresponding flow record. In addition, when a labeled process initiates an outgoing connection, the Egress Monitor will apply that process's label to the outgoing socket and flow. By doing so, it can identify the incoming flow associated with the sending process, connect it with the outgoing flow on the current socket, and establish a concrete relationship between the incoming and outgoing flows. The Egress Monitor then carries out flow-characteristics comparison and endpoint analysis steps, filtering out legitimate lateral traffic and alerting on the remaining pivoting flow pairs.

\textbf{Workflow.} Immediately after a packet is received at the host machine from its network interface card (Rx NIC, Fig.~\ref{fig:full_sys}), the Ingress Monitor extracts the flow information and updates the appropriate flow record (\textit{flowStats}, Fig.~\ref{fig:full_sys}). Using the incoming flow ID as a label, it then labels the socket used by the flow in order to transfer the flow ID from there to the process that accesses the receiving socket (\textit{socketIn}, Fig.~\ref{fig:full_sys}). Once a process accesses a labeled socket, the Process Monitor takes over and tracks it for the remainder of its lifespan. For example, if a labeled process is cloned, the Process Monitor transfers the label to the resulting new process (\textit{tgids}, Fig.~\ref{fig:full_sys}). If a labeled process initiates an outgoing connection, the Egress Monitor will apply the process's label to its new outgoing socket (\textit{socketOut}, Fig.~\ref{fig:full_sys}). The Egress Monitor then intercepts outgoing packets from this connection and creates a corresponding flow record to monitor that flow's size and time characteristics (\textit{flowStats}, Fig.~\ref{fig:full_sys}). Using the flow ID that has been propagated to this outgoing connection and stored as a label in \textit{socketOut}, the Egress Monitor can identify the related incoming flow in the \textit{flowStats} map and, on a per-packet basis, compare it to the current outgoing one to identify pivoting characteristics. Finally, the Egress Monitor can check the flow's destination and, depending on the detection outcome, either forward the packet as usual or send an alert to userspace.

\subsection{Information Flow Tracing}
\label{subsec:info_flow_tracing}
We present below the operation of each kernel-level module as they coordinate to identify inter-flow relationships and collect the information required to detect pivoting.

\subsubsection{Network Ingress Monitor} 
This component monitors incoming flows and bridges the gap between raw packets and their receiving processes. It allows \sysname{} to effectively monitor information flows immediately upon their arrival.

\textbf{Monitoring Incoming Flows.} First, incoming flows are monitored on a per-packet basis at the earliest possible point in the kernel using XDP, which executes immediately after the NIC has copied a raw packet into the host kernel. When a flow arrives, its characteristics are stored in the \textit{flowStats} eBPF hash-map as key-value pairs using the flow ID $F_1$ (consisting of the five-tuple [Source IP, Destination IP, Source Port, Destination Port, Protocol]) as the key and the flow state $(t,s,d)$ ([arrival\_time, flow\_size, direction]) as the value. Flow direction is established based on the direction of its first packet, and flow\_size is updated on a per-packet basis. These records are accessible to the other components of \sysname{}, later used to conduct the comparisons required to identify pivoting.

\begin{figure}
    \centering
    \includegraphics[width=0.8\linewidth]{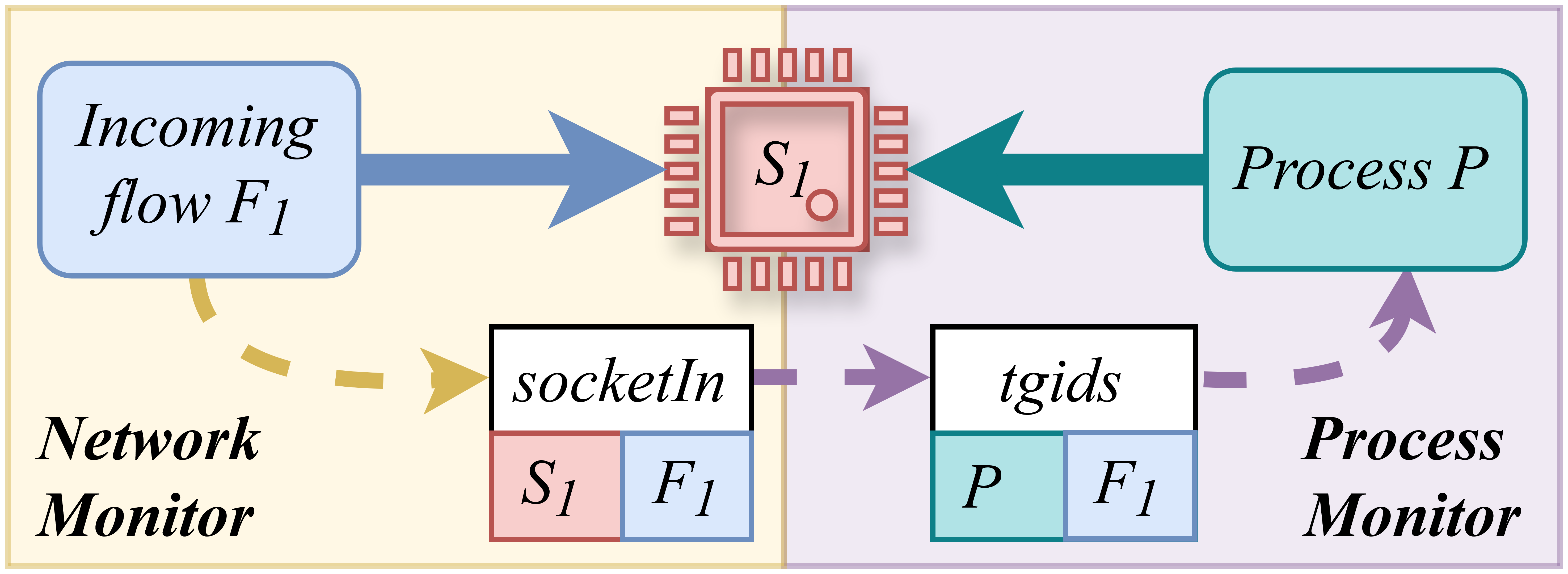}
    \caption{The Network Ingress Monitor labels the socket $S_1$ so that the Process Monitor can associate the incoming flow ID $F_1$ with its receiving process.}
    \label{fig:sk_2_proc}
\end{figure}

\textbf{Connecting Flows to Their Receiving Processes.} In order to track the flow of information across the machine, we must label processes in such a way that we can identify which incoming network flows belong to them. Using the five-tuple flow ID, $F_1$, as a label, \sysname{} can refer back to the \textit{flowStats} map and identify characteristics of incoming network flows. However, rather than receiving raw packets, processes typically receive data via network sockets, preventing the Network Ingress Monitor from locating the destination process directly from the raw packets used to monitor incoming flows. To address this, incoming flows are instead associated with their recipient socket (Fig.~\ref{fig:sk_2_proc}): the Network Ingress Monitor uses the \textit{socketIn} map to store socket IDs ($S_1$) with the corresponding incoming flow ID ($F_1$). This way, when a process accesses a socket, the Process Monitor can look up that socket in the \textit{socketIn} map and identify the corresponding flow.

\subsubsection{Process Monitor}
Now that the Process Monitor has access to both $F_1$ (via the \textit{socketIn} map) and the receiving process's thread-group ID (TGID) via the \textit{accept} system call, it can label the process with the appropriate incoming flow ID. Process labeling is achieved by creating a key-value pair in the \textit{tgids} map, where the key is the process's TGID, $P_1$, and the value is the flow ID $F_1$. The TGID is used instead of the process ID (PID) to accomplish process-level tracing as described in Section \ref{subsec:chals}. The \textit{tgids} map is then used to track the life cycle of labeled processes. Specifically, every time the \textit{execve}, \textit{clone}, \textit{fork}, or \textit{exit} system calls are invoked, the Process Monitor uses a corresponding eBPF program to check whether or not the calling process has an entry in the \textit{tgids} map and react accordingly. On a call to \textit{execve}, \textit{clone} or \textit{fork} from a labeled process, the Process Monitor creates a new entry in \textit{tgids} with the TGID of the new process as the key and the flow ID associated with the calling process as the associated value (Fig.~\ref{fig:full_sys}). Labels are thus passed on from parent to child and \sysname{} can continue to infer causality between flows in the face of complex process call-chains. On calls to \textit{exit}, the exiting process is removed from \textit{tgids}.

\subsubsection{Network Egress Monitor}
Finally, the Network Egress Monitor identifies outgoing connections, monitors outgoing flows, and detects causal relationships between incoming and outgoing flows.

\textbf{Identifying Outgoing Connections.} First, the Network Egress Monitor must identify which processes, if any, initiate outgoing connections. This is done by monitoring the creation of outgoing sockets. Specifically, when a process creates an outgoing connection with a new socket, the Network Egress Monitor checks the TGID of the process against entries in \textit{tgids} to see if it has a label. If the process is labeled, a workflow similar to the Network Ingress must be used to bridge the gap between the process at the socket and the corresponding raw packet data that will be used to monitor outgoing flows. To this end, the Network Egress Monitor uses the \textit{socketOut} map to copy the label from the process to the socket. \textit{socketOut}, like \textit{socketIn}, can then bridge the gap between processes and outgoing flows. 

\textbf{Monitoring Outgoing Flows.} The Network Egress Monitor monitors outgoing flows. It shares the same \textit{flowStats} map with the Network Ingress Monitor, creating outgoing flow records and updating the flow size on a per-packet basis. A shared map for incoming and outgoing flows is important because network flows are \textit{bidirectional}, meaning that an incoming flow will have both incoming and outgoing packets. Therefore, when the Network Egress Monitor encounters an outgoing packet from an existing, incoming flow, it can directly update the corresponding record rather than checking separate maps to verify if the flow exists.

\textbf{Identifying Causal Relationships.} Finally, by using the information gathered across the host system, the Network Egress Monitor is able to establish causality between two flows. For every outgoing flow, the Network Egress Monitor checks the \textit{socketOut} map to see if it is associated with a labeled socket. If an entry exists for the current socket, then a relationship exists between an incoming flow and the current outgoing one, and we establish a causal relationship between the two. Because the label applied to processes and sockets consists of the incoming flow ID, the corresponding record of that incoming flow can be directly accessed in the \textit{flowStats} map and used to decide if this causal relationship exhibits pivoting characteristics.

\textbf{Storage Management.} \label{subsec:storage_management} To maintain small data stores and avoid draining host resources, each component of \sysname{} implements eBPF \textit{Least-Recently-Used (LRU)} hash maps. Slightly different than first-in-first-out (FIFO), when a hash map reaches maximum capacity, the LRU approach evicts the least recently been accessed (for any reason, be it write, modification, or read) entry rather than the oldest one (as in FIFO). Because old flows are not necessarily inactive, this approach ensures that \textit{active flows} being monitored by \sysname{} are less likely to be evicted than dormant or complete ones, whose map entries will not have been recently accessed. eBPF provides native support for LRU hash maps \cite{bpf_maps}, allowing \sysname{} to offload map access tracking to the eBPF subsystem.

\sysname{} can perform access-based eviction due to the insight that pivoting flows \textit{maintain symmetry for the duration of their lifetime} (confirmed in Section \ref{sec:evaluation}). For example: when a pivoting flow pair becomes inactive, both the incoming and outgoing will cease activity as they're propagating identical content (per the \textit{active pivoting} described in Section \ref{subsec:threat_model}). If a flow pair is evicted before completion, \sysname{} will simply re-add the required entries for each flow upon reception or delivery of the constituent flows' packets, resetting the timing and size and maintaining their relationship. This way, \sysname{} reduces storage requirements by using small map sizes with high turnover and reduces instrumentation by relying on the kernel-native eBPF subsystem.

\subsection{Identifying Pivoting}
\label{subsec:id_pivoting}
Once \sysname{} identifies a causal relationship between two flows via process tracing, flow characteristics comparison and endpoint frequency analysis steps are performed to filter out pivot attack traffic from other tunneling activity.
 
\textbf{Flow Characteristics.} First, \sysname{} checks whether the connected flows' \textit{arrival times} and \textit{sizes} fall within the given time and size windows. The time window is the maximum allowable difference between the start times of two flows within which they could be considered to be active pivoting. For example, if we consider two flows, flow A and flow B, and the time window is one second, then $abs(A_{start} - B_{start}) \le 1$s would indicate that pivoting may be occurring. The size window is the maximum allowable difference between the sizes of two flows within which they would be considered pivoting: if the size window is 6kB, then $abs(sizeof(A) - sizeof(B)) \le 6$kB indicates a pivoting flow. If both conditions are satisfied, \sysname{} proceeds to the endpoint analysis step. If at least one of these conditions is not met, the flows are not considered to be pivoting and the endpoint analysis is not performed.

\textbf{Endpoint Connection Analysis.}\label{subsec:ep_conn_analysis} Having identified pivoting, \sysname{} will check the destination of the outgoing flow to further evaluate its trustworthiness. Flows traveling to frequently visited endpoints are considered more trustworthy than those traveling to infrequently visited ones. Specifically, if the number of visits to a certain endpoint \textit{E} (expressed as an IP and destination-port pair) exceeds an administrator defined threshold, \sysname{} will ignore the pivoting activity as it is considered non-anomalous. Conversely, if the number of visits to endpoint \textit{E} are \textit{below} the administrator defined threshold, \sysname{} will generate an alert and send it to userspace.

In order to track the number of connections per endpoint, \sysname{} uses a hash map, \textit{endpoints}, to store the key pair [Destination IP, Destination Port] alongside a corresponding counter value. For every new outgoing flow, the Egress Monitor will check if a pair exists in \textit{endpoints} with the flow's current information. If yes, it will increment the counter value stored with the current flow's pair. If no entry exists for the pair, it will create a new entry to begin tracking the new endpoint. In order to avoid data overflow from infinitely increasing counters, we perform min-max scaling across all counters to a range of [0, \textit{threshold}] when any single counter reaches 90\% of its capacity ($\sim3.86*10^9$ as unsigned, 32-bit integers). This strategy reduces counter values while maintaining the distribution of connections across tracked endpoints, allowing \sysname{} to avoid overflows while continuing to identify anomalous connections.

\section{Evaluation}
\label{sec:evaluation}
We evaluate \sysname{} by conducting experiments in both a controlled testbed and in two separate live network deployments. This section presents the implementation, setup details, and results of our evaluations in each environment.

\subsection{Implementation}
We implement a prototype of \sysname{} with around 450 lines of C code for the kernel components and 100 lines of Python for the userspace agent.

To evaluate the benefits of \sysname{} against a purely flow-characteristics-based one, we implement a solution that performs detection using only flow time and size characteristics, which we refer to as the FCB (flow-characteristics based) solution. Implemented as a host-based eBPF system, it tracks incoming and outgoing flows on a per-packet basis. We isolate this FCB solution because it forms the detection foundation of both the in-network system \spotlight{} \cite{spotlight} and the host-based \textit{APIVADS} \cite{apivads}: despite differing in deployment and architecture, both ultimately identify pivoting via flow timing and size analysis.

\subsection{Tuning Detection Parameters}
Different parameter values (time-window, size-window, and endpoint connection threshold) will affect the performance of \sysname{}. Optimal values would result in no missed pivoting attacks and a negligible number of false positives requiring confirmation from an administrator. 

Pivoting flows propagating the same data will be identical in size, barring any differences in the protocol or other control mechanisms \cite{spotlight}. Such control mechanisms include, most significantly, connection setup: since an incoming flow must initialize the connection with the pivot before propagating data to the target, it will naturally be some number of bytes larger than the outgoing flow. Analysis of common pivoting tools \emph{SoCat} \cite{socat}, \emph{Chisel} \cite{chisel}, \emph{SSH} \cite{ssh}, and \emph{Nmap} \cite{nmap} reveals connection setup sizes ranging from $<$50B in the case of SoCat to around 5500B in the case of Chisel, while setup times range between 100ms (SoCat) and 600ms (Chisel).

Taking advantage of the characteristics of the four attacks, which have been identified in real-world scenarios \cite{afr_fin, financial_atks, pysa, nmap1, nmap2} and represent a range of commonly identified activities in APT attacks \cite{mitre_atk, apt_survey}, we present a baseline set of parameters, $S_w = 6500$B and $T_w = 1$s, confirmed in Section~\ref{subsec:dectection}. These parameters remain narrow enough to filter out false positive results while still capturing the characteristics of the given attacks with room for natural variations introduced by network conditions. When carrying out endpoint frequency analysis, we adopt a baseline threshold of 10 connections before an endpoint is considered common, placing our threshold at the upper end of the effective daily range (5-11) identified by \textit{Hopper} \cite{hopper}. While administrators may not be able to predict the exact tools and configurations used by potential attackers against their network, these baseline parameters provide a solid foundation based on common real-world pivoting tools and established detection insights \cite{hopper, distdet}.

\subsection{Testbed Environment} 
\textbf{Setup.} Our testbed environment consisted of three interconnected servers with one acting as the pivot, another acting as a target, and the third hosting a 24-node virtual network that contained both targets and attackers. The server hosting the virtual network contained an Intel® Core™ i7-9700 CPU @ 3.00GHz, 16GB RAM, and a 40Gbps Agilio® LX SmartNIC while the others each contained an Intel® Xeon® Silver 4210R @ 2.40GHz CPU, 32GB RAM, and a Mellanox Bluefield 1 SmartNIC. The operating system on all testbed servers was Ubunutu 24.04.4 with kernel version 6.8.

\textbf{Network Traffic.} In this scenario, Selenium \cite{selenium} was used to generate synthetic web traffic mimicking user behaviour when interacting with a PHP-based web server that retrieves and serves additional resources from the external Internet. This traffic is modeled after the Canadian Institute of Cybersecurity's Intrusion Detection System 2017 (CIC-IDS2017) dataset \cite{ids2017}, specifically from the perspective of the web servers contained therein.

\textbf{Pivot Attacks.} We implement four pivot attacks to cover three important categories of activity that can occur during APT attacks: reconnaissance, malware propagation, and data exfiltration \cite{apt_survey}. These categories are commonly found in APT attacks and are listed in the MITRE ATT\&CK matrix for enterprise networks \cite{mitre_atk}. Attacks are performed using the following: \emph{SoCat} \cite{socat}, due to its prominence in previous pivoting works \cite{spotlight, p4control, pivotwall}; \emph{Chisel} \cite{chisel}, as it has been present in several recent cyber attacks \cite{afr_fin, financial_atks, pysa};  \emph{SSH} \cite{ssh}, which has also been present in many previous works \cite{apivads, spotlight, hopper}; and \emph{Nmap} \cite{nmap}, a common tool for reconnaissance that has been reported in recent real-world attacks \cite{nmap1, nmap2}. 
Attacks are initiated from both outside and inside the virtual subnet to emulate both external attacks and attack traffic already traversing the network.

\subsection{Testbed Results}
\label{subsec:dectection}
We generate background web traffic as previously described and inject attacks against all network nodes via an intermediary web server, i.e., the pivot node, to evaluate the performance of \sysname{} on a per-attack basis.
We compare \sysname{}'s overall performance against an FCB solution. We find that our detection approach reduces the average FPR from an average of 42.38\% to an average 0.18\% while improving the detection rate by an average 31.49\% across all size and time windows.

\begin{figure}[ht]
\begin{subfigure}{\linewidth}
    \centering
    \includegraphics[width=\linewidth]{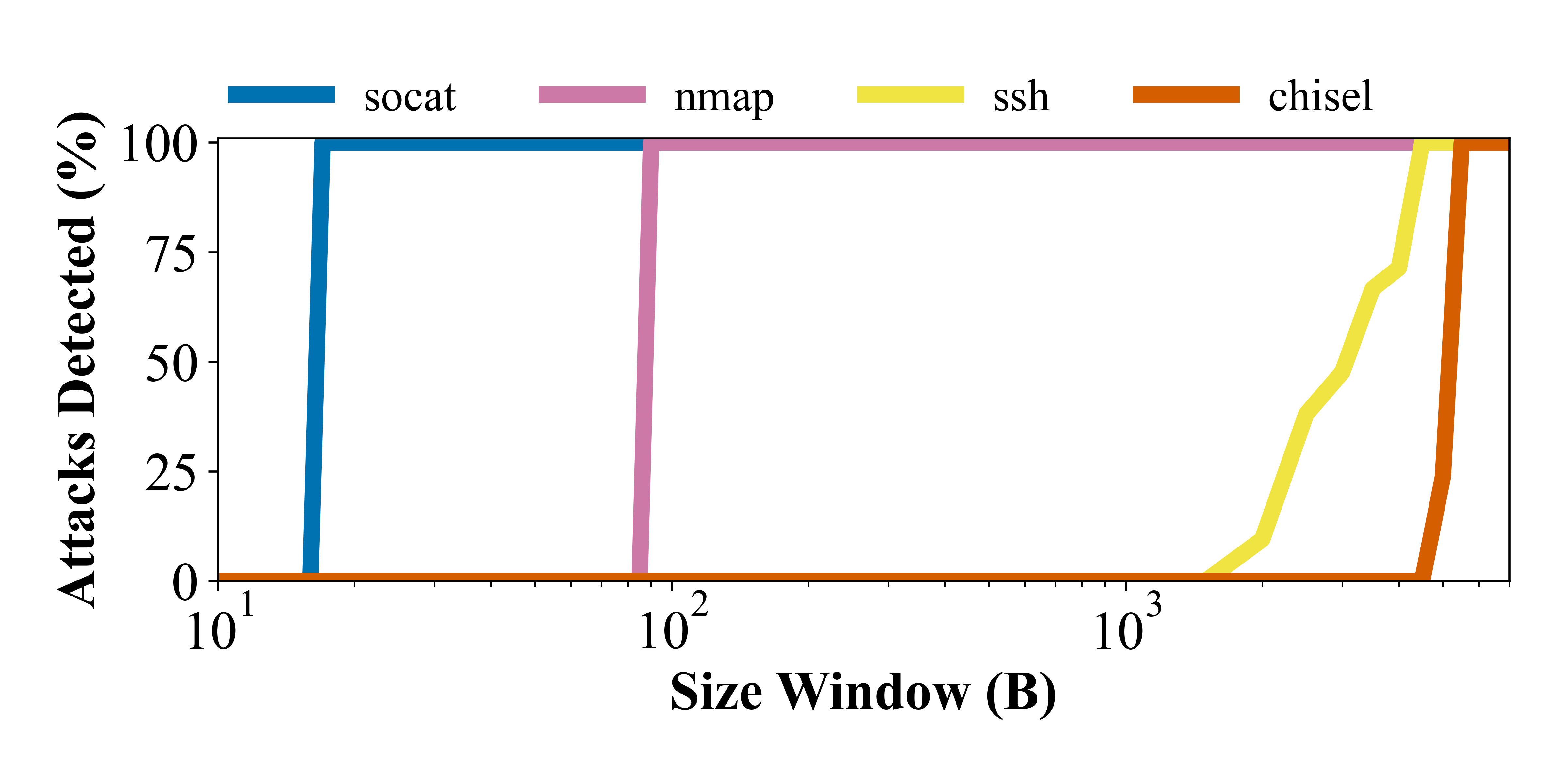}
    \caption{Varying size-window.}
    \label{fig:per_attack_det_s}
\end{subfigure}

\begin{subfigure}{\linewidth}
    \centering
    \includegraphics[width=\linewidth]{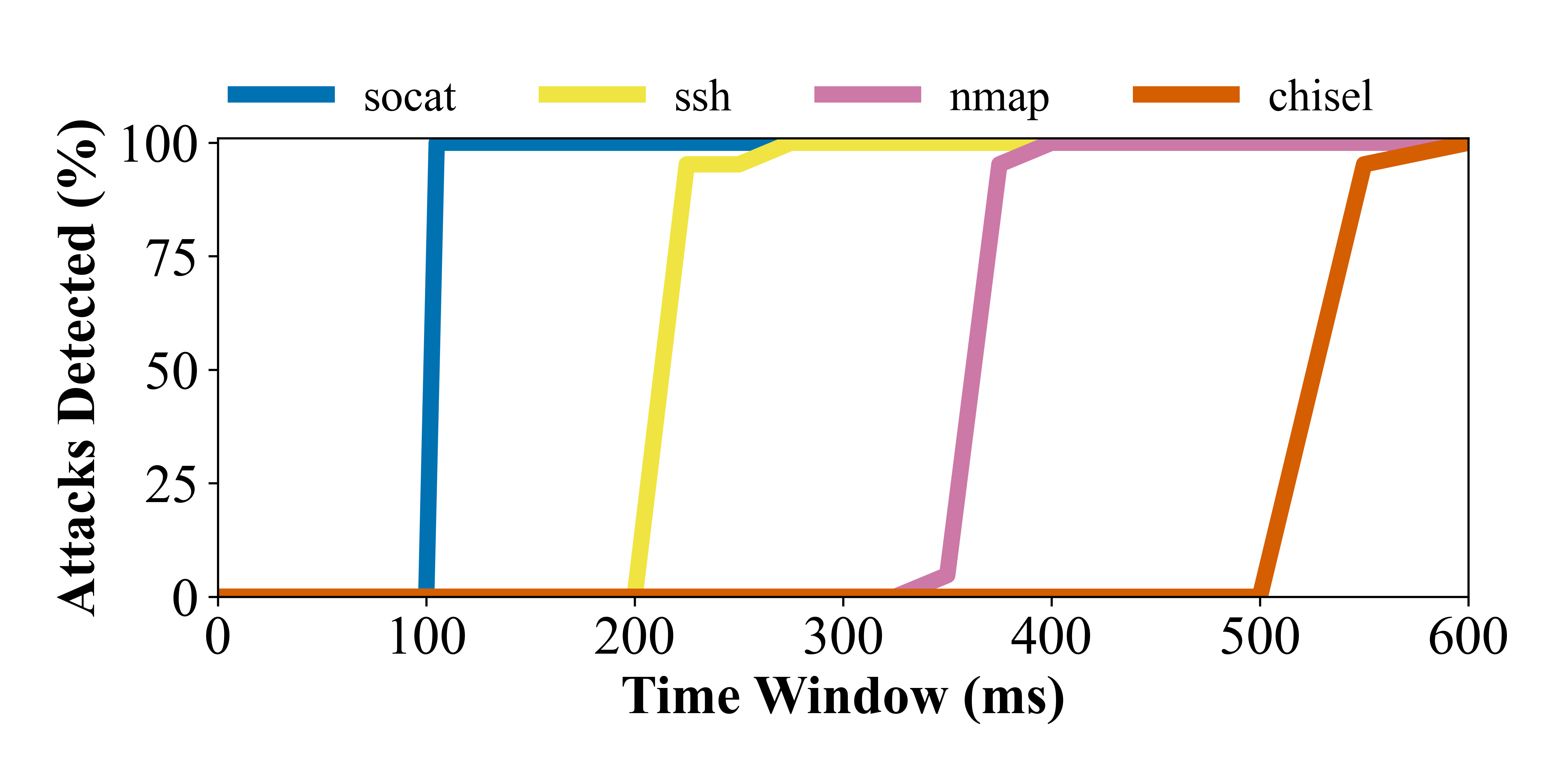}
    \caption{Varying time-window.}
    \label{fig:per_attack_det_t}
\end{subfigure}
\caption{Effects of parameter tuning on attack detection in \sysname{}. Detection plateaus at 100\% for every tool almost immediately after encapsulating its connection-setup thresholds. SSH displays variable sizes during setup, leading to a more gradual approach to its plateau.}
\label{fig:per_attack_det}
\end{figure}

\textbf{Size Window Detection Effects.} We lock the time window at one second and the endpoint connection threshold at 10 while varying the size window to assess its effects on detection accuracy when considering different attack tools. The results in Fig.~\ref{fig:per_attack_det_s} show that the true-positive detection accuracy increases along with the window size in all four attacks, reaching a plateau at a different threshold for each. Because these tools must establish a connection between the pivot node and the attacker node \textit{before} transmissions begin to the target node, initial connection setup is not forwarded across both the pivot's incoming and outgoing flows, and the incoming flow will be slightly larger than the outgoing. This initial connection size varies across tools, and we observe different minimum size thresholds for each attack. Socat and Nmap show small connection setup sizes: \sysname{} identified each with 100\% success using $S_w=17$B and $S_w=90$B, respectively. SSH and Chisel show relatively larger setup sizes, requiring $S_w=4500$B for the former and $S_w=5500$B for the latter to achieve 100\% detection accuracy. Additionally, SSH has a more gradual slope towards 100\% detection than the other tools. This shows that SSH has more variability in its setup size, likely due to the variety of possible options and parameters allowed by the protocol. Once the size window is large enough to encapsulate a certain attack, it has no further effects on true-positive detection.

\textbf{Time Window Detection Effects.} We lock the size window at 6500B and the endpoint connection threshold at 10 while varying the time window to observe its effects on a per-attack basis. With results shown in Fig.~\ref{fig:per_attack_det_t}, we can see that the time window is reduced to the order of milliseconds before attacks begin to be missed. This parameter must encapsulate the clock time required for a pivoting tool to create its initial connection to the pivot node and instigate the outgoing flow. Similarly to the size window, effective time windows will be determined by the characteristics of the tool used, but will additionally be influenced by the environment in which the attack is performed due to variable network and system latencies. Socat again shows the minimal requirement at only $T_w=105$ms required for 100\% detection, with Chisel requiring the largest window for 100\% detection success at $T_w=600$ms. SSH and Nmap require at least $T_w=275$ms and $T_w=400$ms, respectively. 

\begin{figure}
\begin{subfigure}{.25\textwidth}
  \centering
  \includegraphics[width=\linewidth]{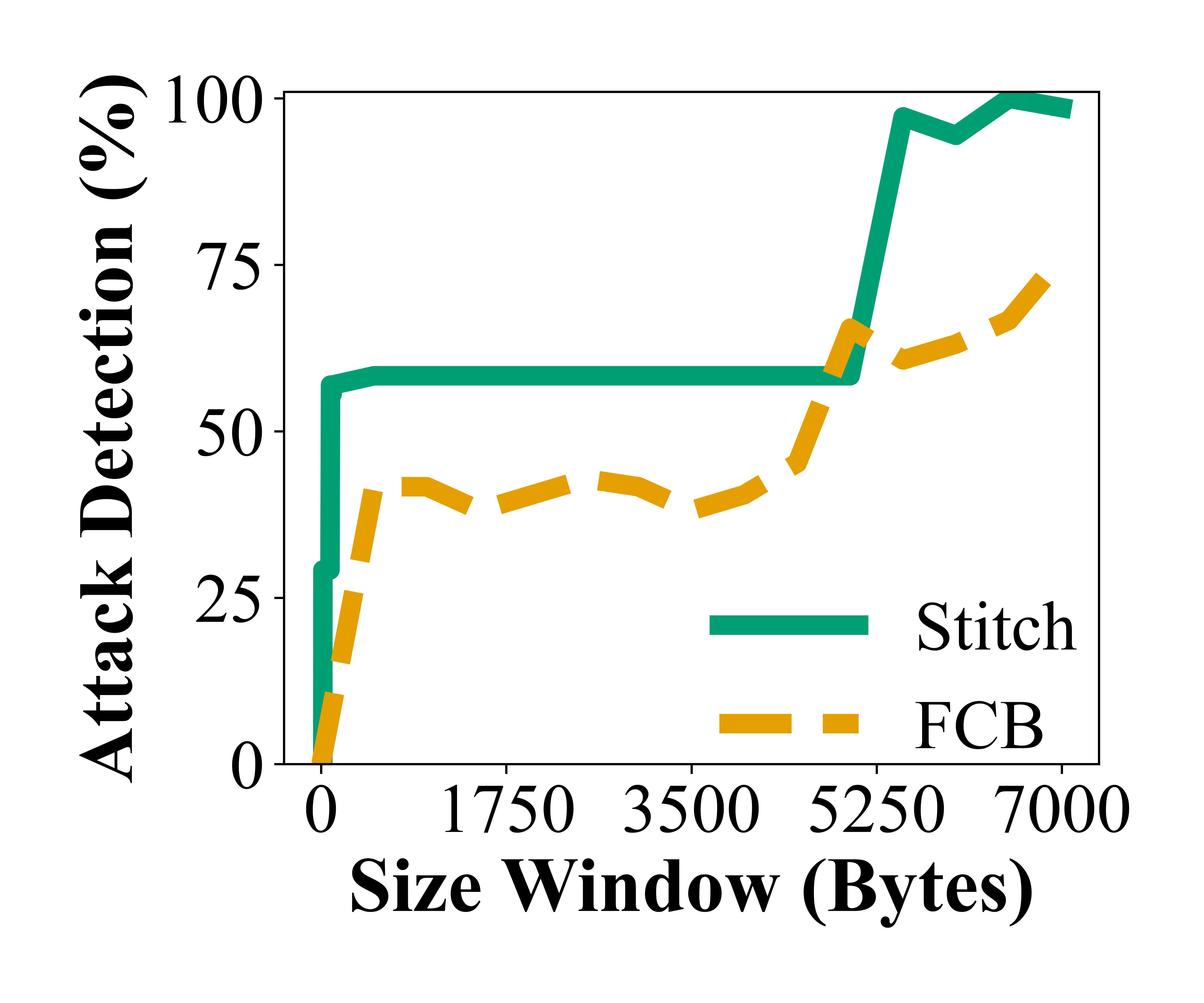}
    \caption{Varying size-window.}
  \label{fig:det_acc_s}
\end{subfigure}%
\begin{subfigure}{.25\textwidth}
  \centering
  \includegraphics[width=\linewidth]{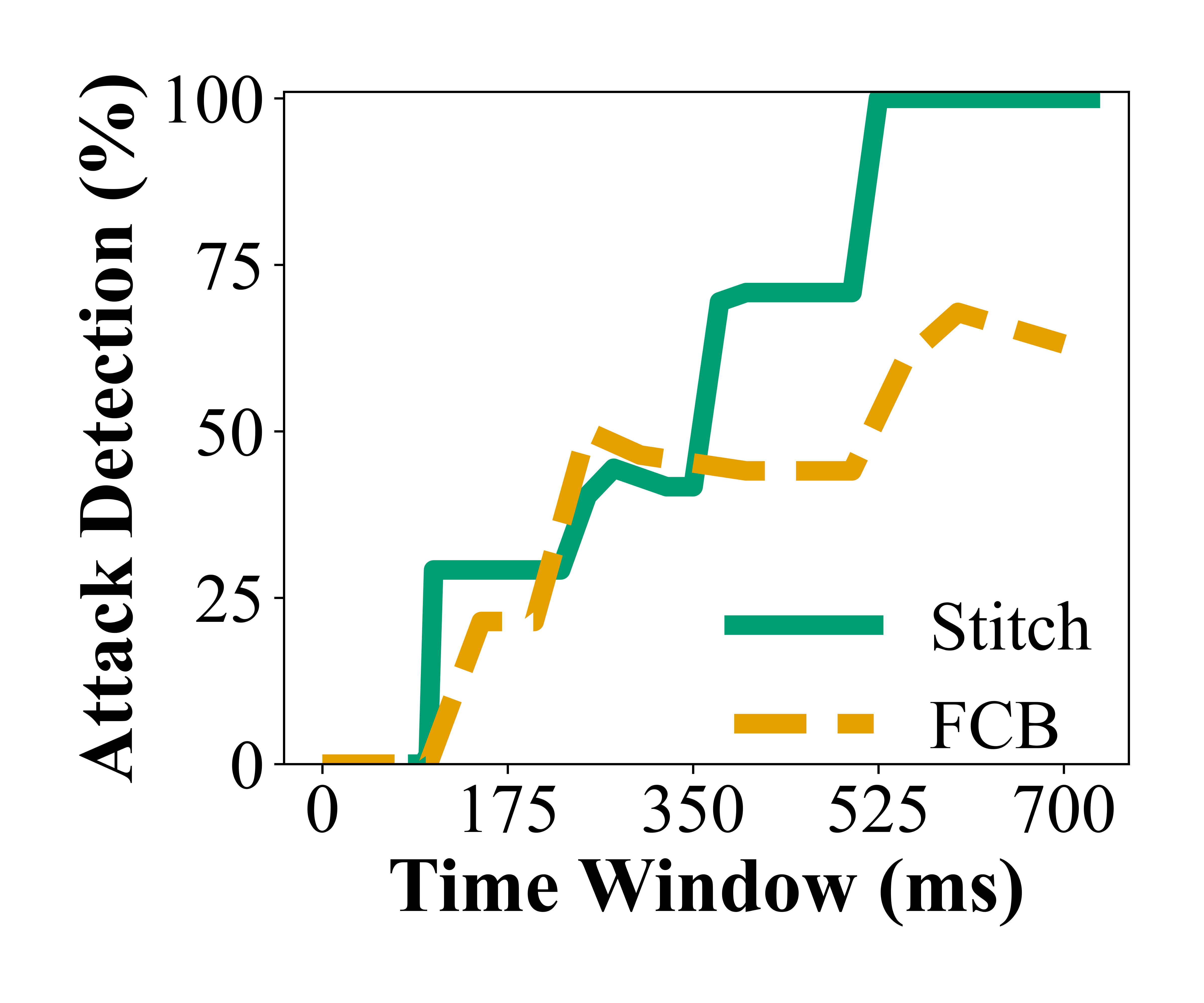}
    \caption{Varying time-window.}
  \label{fig:det_acc_t}
\end{subfigure}
\caption{Overall attack detection rates of \sysname{} and the FCB solution. Superfluous consideration of unrelated flows causes erratic detection in purely FCB solutions.}
\label{fig:det_acc}
\end{figure}

\textbf{Overall Detection Accuracy.} True positive detections are presented in Fig.~\ref{fig:det_acc}. The endpoint connection threshold set to any value $>$1 had no affect on attack detection in this testbed scenario. Results in Fig.~\ref{fig:det_acc} show that the FCB solution can maintain a similar detection rate to that of \sysname{} when using more restrictive parameters (e.g., $T_w \le 350$ms), but struggles to keep up when the parameters are widened, trailing \sysname{} by an average 31.49\% across all size and time windows. Additionally, the FCB solution only achieved a maximum detection rate of 76.19\% for this attack set. We identify that it struggles in particular to detect SSH and Chisel attacks, likely due to the tools' relatively large and slow initial setups (4500B/275ms, and 5500B/600ms, respectively). FCB systems use arrival time to group potentially related flows: any incoming flows arriving within the specified time window before the outgoing one are added to this group, evicting the next oldest entry. Because SSH and Chisel require wider time windows, their instigating flows are likely evicted from the group before being able to reach the required size (compared to the outgoing) in order to be flagged. This scheme also produces variable detection behaviour: relying on chance to group flows can result in the erratic success rate shown in Fig.~\ref{fig:det_acc_s}. By first establishing causality via IFC, the detection scheme in \sysname{} removes the chance element inherent to timing-based flow grouping and increases the reliability of pivot detection.

\begin{figure}
  \centering
  \includegraphics[width=0.75\linewidth]{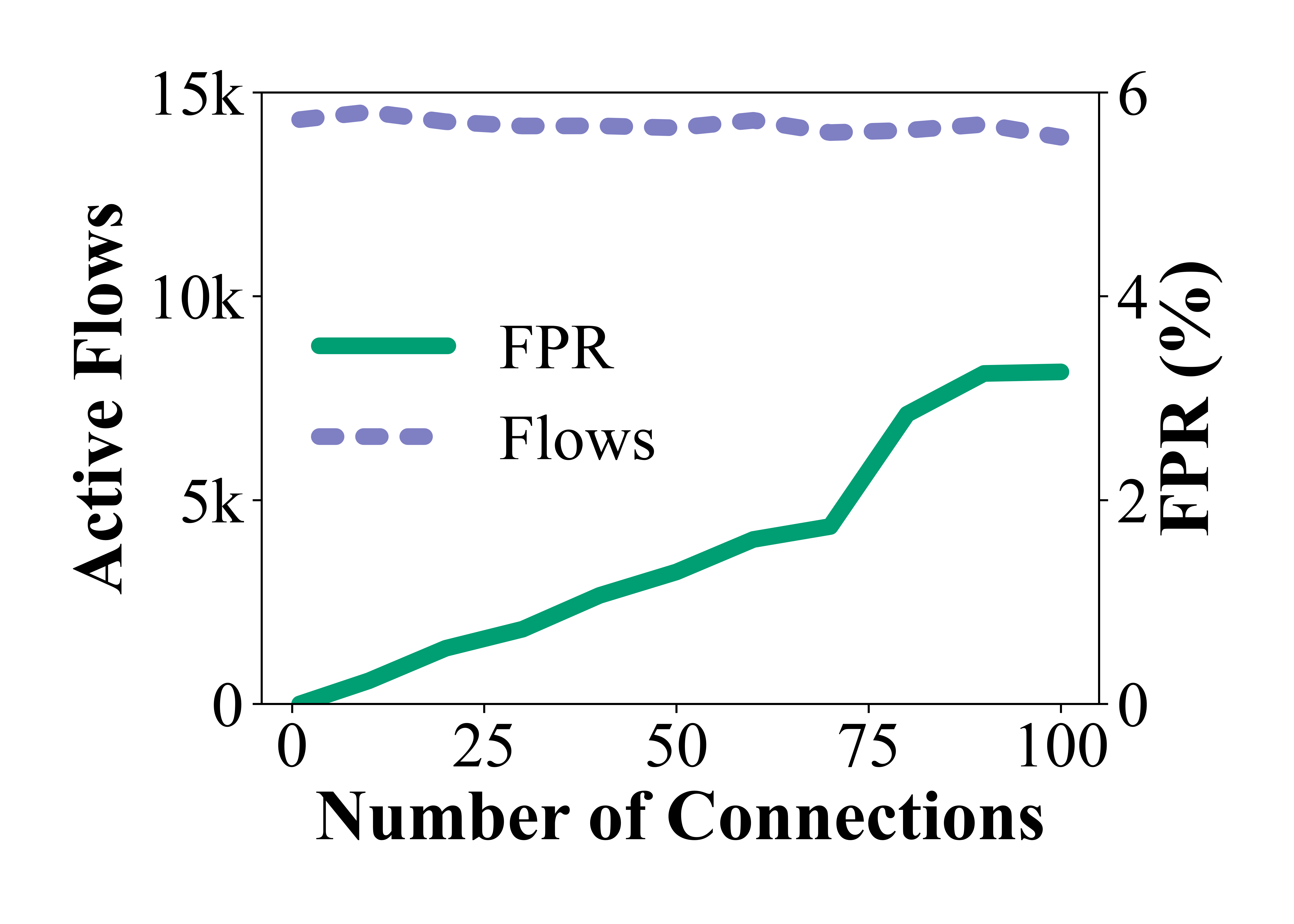}
\caption{Effects of different connection counting thresholds (Number of Connections) on FPR presented with the number of background flows (Active Flows) at the time of capture.}
\label{fig:ep_fpr}
\end{figure}

\begin{figure}[ht]
\begin{subfigure}{0.5\linewidth}
    \centering
    \includegraphics[width=\linewidth]{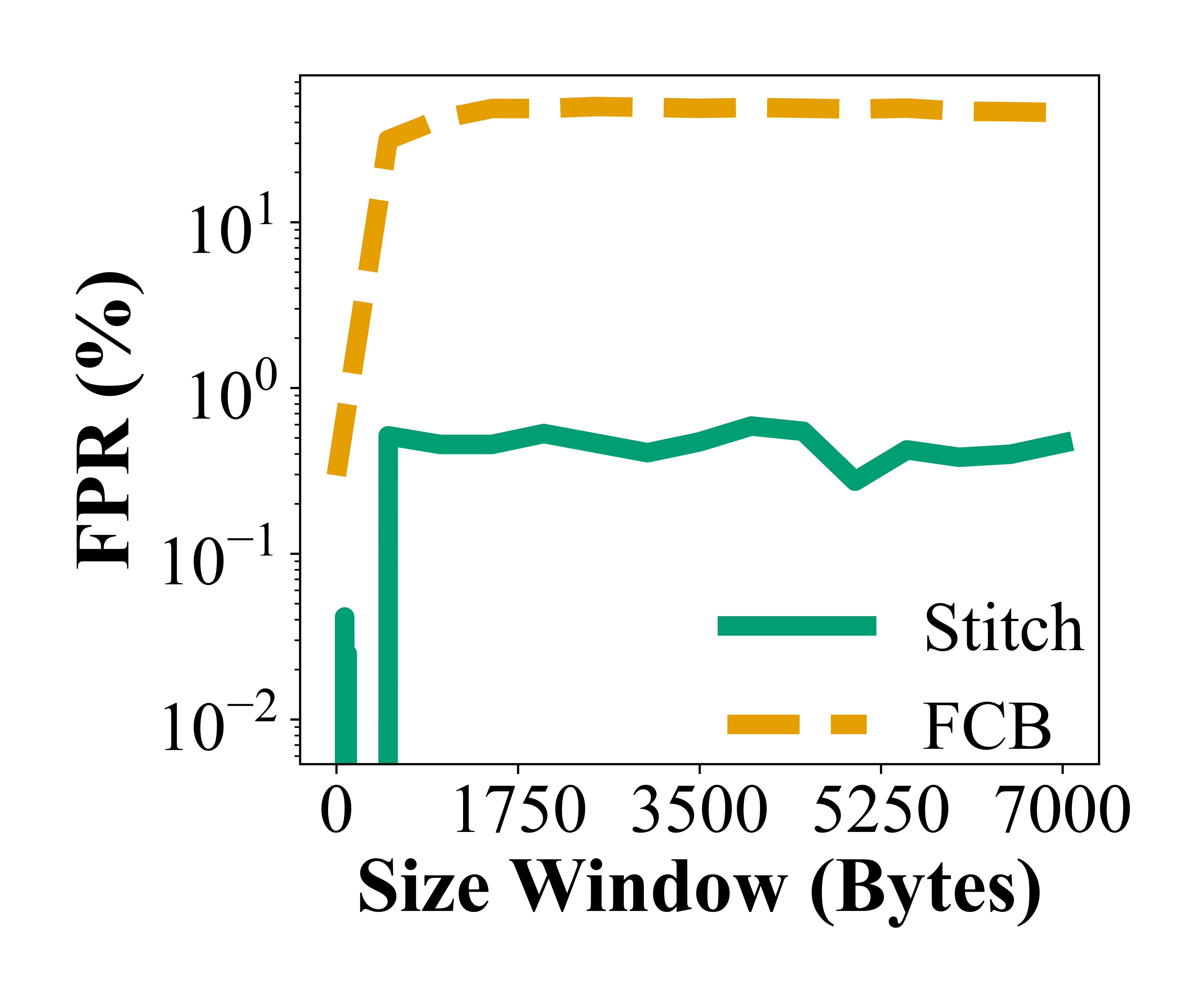}
    \caption{Varying size-window.}
    \label{fig:fps_sz}
\end{subfigure}%
\begin{subfigure}{.5\linewidth}
    \centering
    \includegraphics[width=\linewidth]{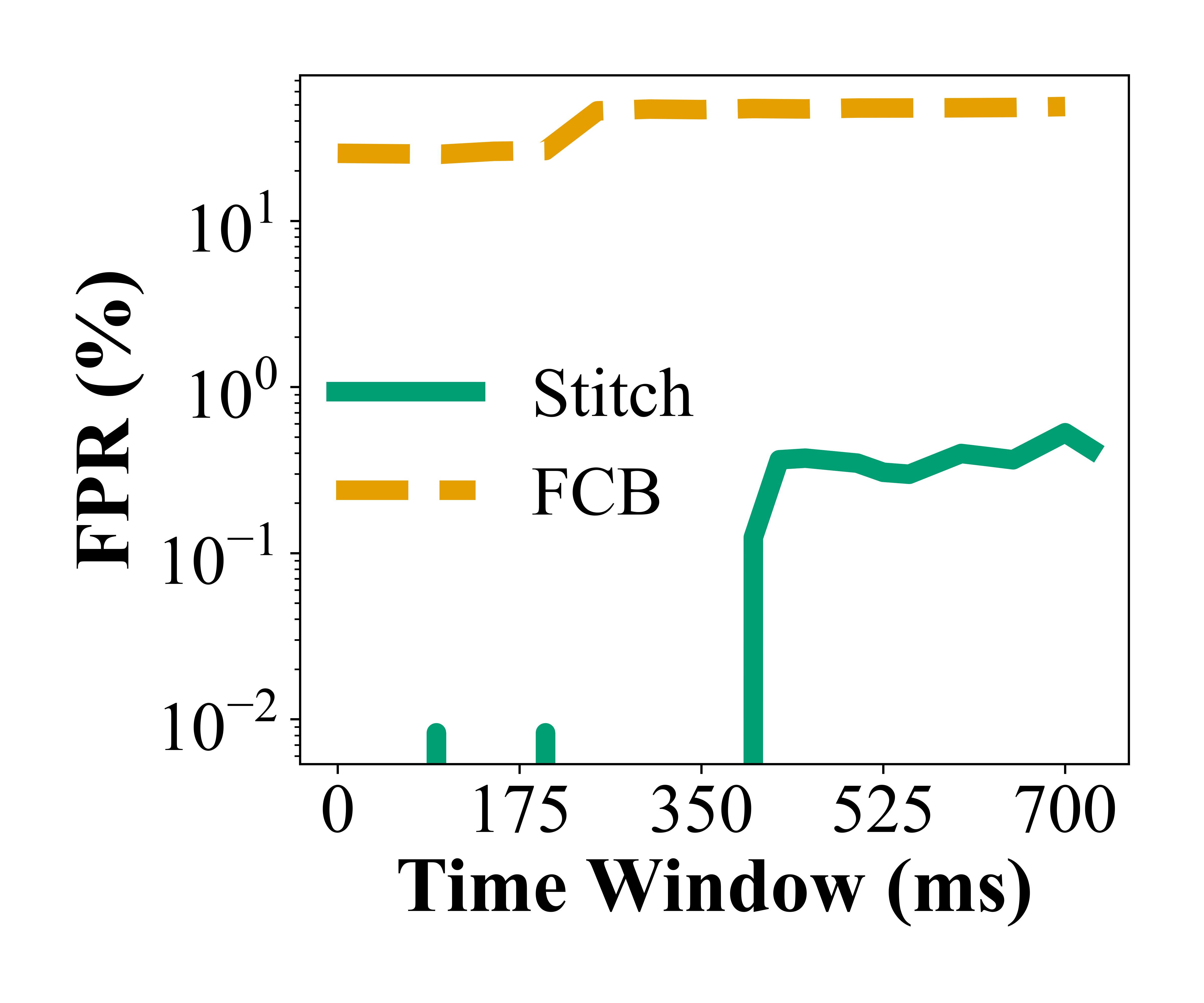}
    \caption{Varying time-window.}
    \label{fig:fps_t}
\end{subfigure}
\caption{FPR of \sysname{} and the FCB solution under different parameters. \sysname{} reduced the FPR from $>$40\% in the case of the purely FCB solution to $<$1\% due to its improved flow association and FPR reduction.}
\label{fig:fps_all}
\end{figure}

False-positive reports in \sysname{} are shown in Fig.~\ref{fig:fps_all} and Fig.~\ref{fig:ep_fpr}. Fig.~\ref{fig:fps_all} shows that false positives remain relatively steady across different size windows and time windows once a certain threshold is reached. However, Fig.~\ref{fig:ep_fpr} shows that the FPR increases linearly over a growing endpoint connection threshold. A threshold of 10 results in an FPR of 0.23\% and a threshold of 100 results in an FPR of 3.26\%.

False-positive reports occur when non-pivoting traffic crosses a node with temporal and size-based similarities. We identify that the external-resource requests made by our testbed web server are of a similar size to the incoming user requests, thus displaying similar characteristics to pivoting traffic. This size difference is encapsulated at minimum by roughly 100B, which means that the FPR plateau in Fig.~\ref{fig:fps_sz} occurs quickly and then remains steady throughout the growing size window. A similar shape occurs in Fig.~\ref{fig:fps_t} with the time window experiments: network conditions can fluctuate and introduce variable latency, but in general the timing of incoming and outgoing web requests is encapsulated by a minimum of 400ms.

Even when encountering such highly symmetric traffic, \sysname{}'s approach reduces the overall average FPR from 42.38\% in the case of the FCB solution to only 0.18\%. The FCB solution faces similar plateaus, but with a far higher number of false positive reports. This occurs because of the non-determinism of timing-based causality inference: although an incoming flow may occur shortly before a similar-sized outgoing one, this provides no guarantee that they are actually related and leads to superfluous consideration of and alerts on unrelated flow pairs. By actively tracing process relationships and monitoring endpoint connections, \sysname{} eliminates the guessing of purely FCB approaches and significantly reduces the FPR. 

\begin{figure}
    \centering
    \includegraphics[width=1\linewidth]{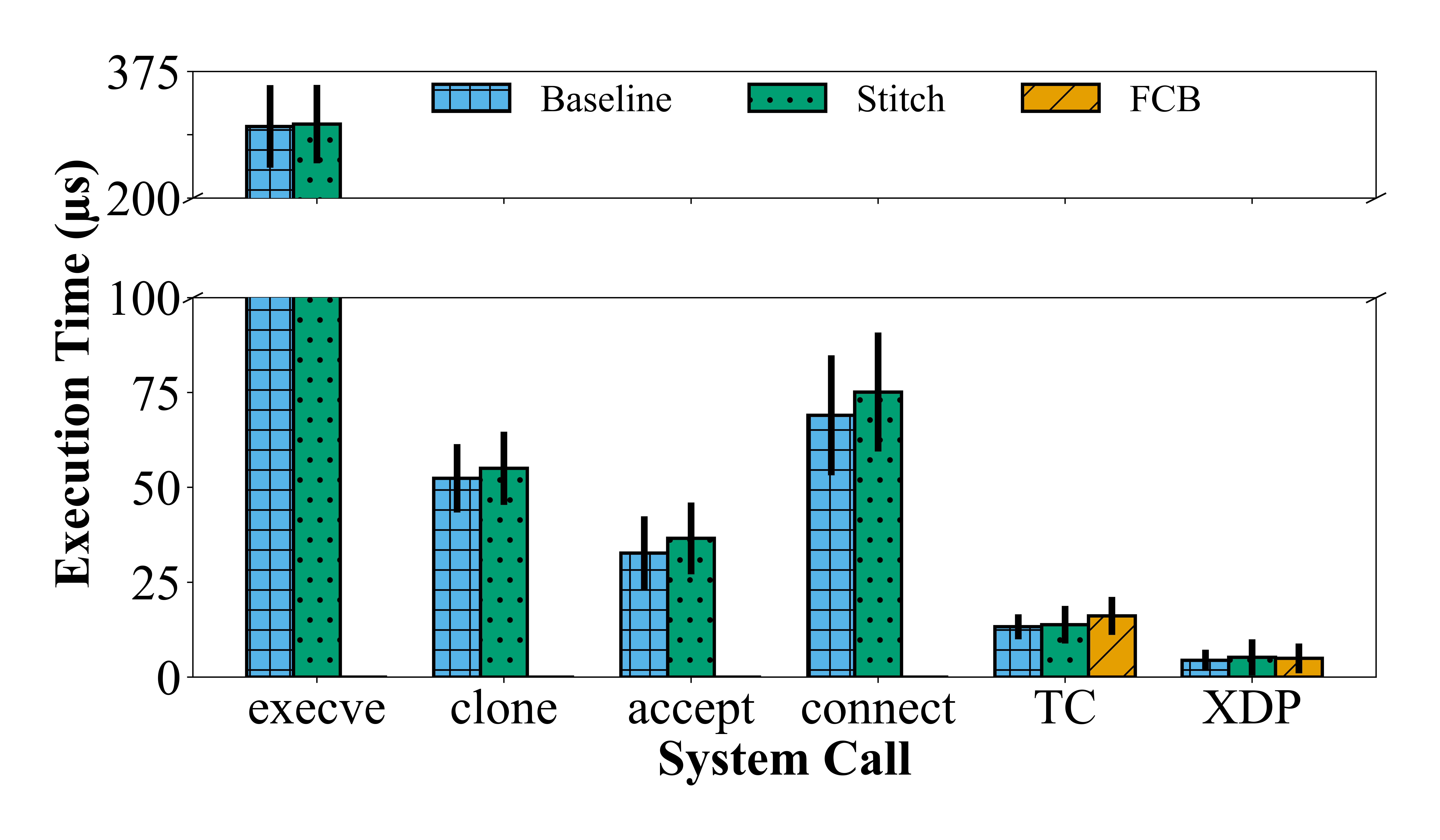}
    \caption{Latency introduced by \sysname{} per system-call, compared to the baseline of the same call without any eBPF and to the FCB solution where applicable. \textit{execve} is much larger than the other calls due to the higher number of tasks performed in the baseline function.}
    \label{fig:lats}
\end{figure}

\textbf{Latency.} We measure the latency introduced by each component of \sysname{} by performing 1000 pivoting attacks with and without each component in place. For each measurement, we compare the total execution time of each system call with \sysname{} attached against its execution time without. These results are shown in Fig.~\ref{fig:lats}.

The process tracing components introduce consistently low overhead, resulting in 1.1\% to 12\% increases in execution times. On a larger call like \textit{execve}, the 3.3us overhead amounts to just over 1\% of the total execution time. The quickest call, \textit{accept}, incurs the largest proportional overhead of 3.9 microseconds, which at this scale is 12\% of the total execution time. The \textit{clone} system call has the largest absolute overhead with an additional 6.1us. 

The latencies introduced by the network hooks of \sysname{} are even lower than those of the process tracing components, introducing 0.6us in the case of TC and 0.8us in the case of XDP. XDP is an eBPF framework designed to perform rapid packet operations. Executed before any costly kernel processing, it is suitable for the high-performance packet processing shown here. No comparable egress framework yet exists, but the TC network hook for eBPF enables similar processing on the egress path.

Because the FCB solution does not perform any process tracing, it requires only the XDP and TC eBPF hooks to perform its detection task. It reduces time for XDP by 0.3us compared to \sysname{}, adding only 0.5us. However, it has an increased TC overhead of 2.86us, which is 3.5x additional overhead compared to \sysname{} for this component. These speeds are due to the fact that, on outgoing flows, the FCB solution must perform causality inference tasks on the outgoing path, whereas \sysname{} pre-establishes causality via IFC. 

\label{subsec:scalability}

\begin{figure}
    \centering
    \includegraphics[width=0.8\linewidth]{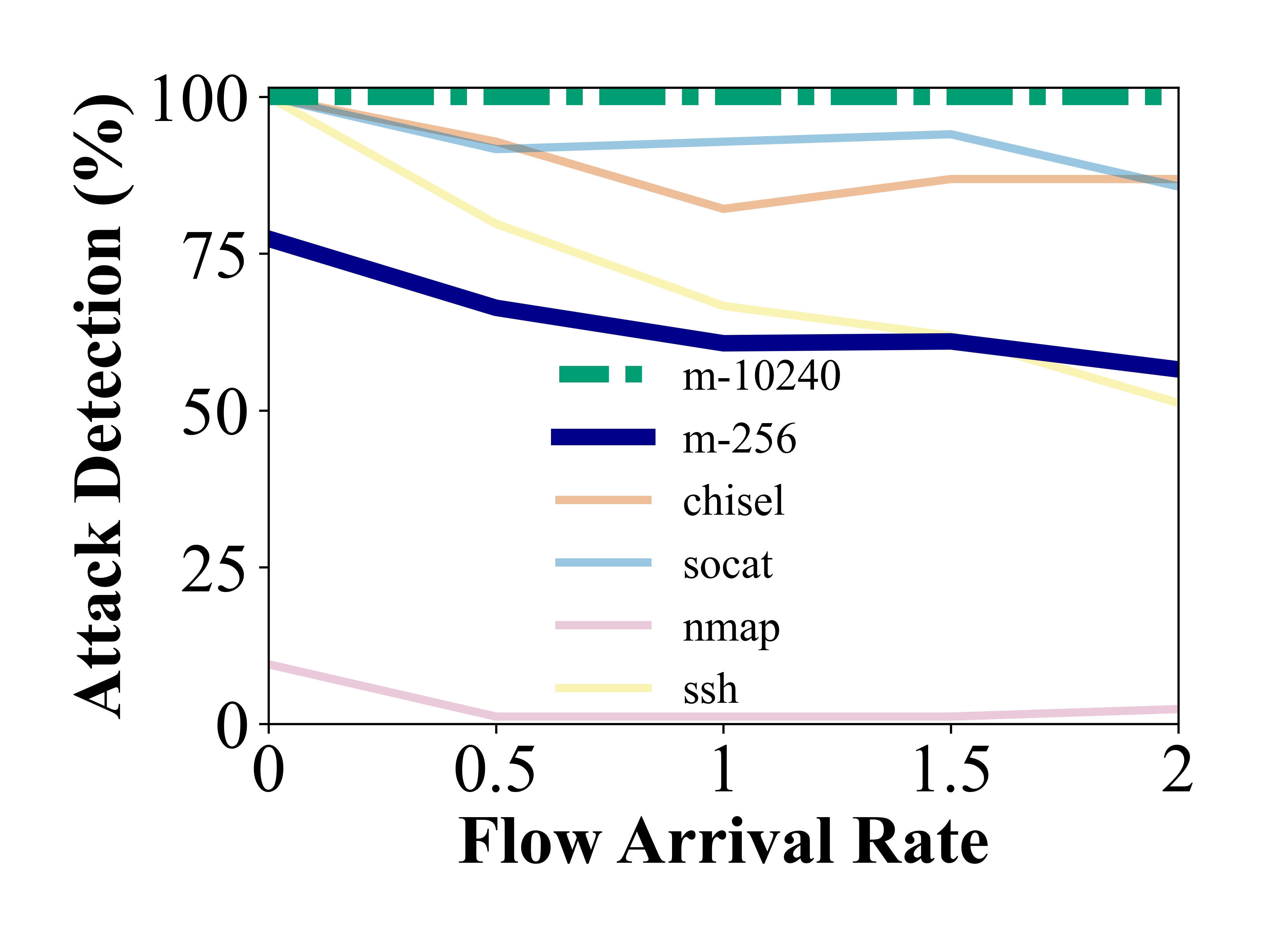}
    \caption{Detection performance of \sysname{} under different traffic loads. Flow Arrival Rate represents the rate of new flows received per second relative to the size (m-256 or m-10240) of the \textit{flowStats} map, which maintains records about currently active flows. E.g., for m-10240 a flow arrival rate of 2 means that \sysname{} is receiving 20480 new flows per second. The per-attack detection breakdown is shown for m-256.}
    \label{fig:det_vol}
\end{figure}

\textbf{Scalability.} Our synthetic traffic, based on CIC-IDS2017, is representative of a mid-sized academic network. In order to demonstrate the behaviour of \sysname{} under a heavier traffic load, we inject pivoting attacks into increasing volumes of background traffic and measure the detection success rate. We use the aforementioned web server as our pivot node and employ \textit{httperf} to flood the server with concurrent connections while attacks are occurring. The size and time windows are tuned to the baseline values $S_w = 6500$B and $T_w=1$s. \sysname{} uses the \textit{flowStats} map to maintain information about ongoing flows, so we use the size of this map as a benchmark for our tests. That is, we send new flows at different rates relative to the size of the \textit{flowStats} map. Using the map's default size of 10240 entries, results in Fig.~\ref{fig:det_vol} show that \sysname{} is able to maintain up to 100\% detection accuracy while handling new flows-per-second (Flow Arrival Rate) at a rate of 2x the size of \textit{flowStats}, which is 20480 new flows per second.

Fig.~\ref{fig:det_vol} also shows how \sysname{} performs overall and on a per-attack basis with a map size of 256 entries. When the size of \textit{flowStats} is very small, \sysname{} cannot maintain records long enough to identify correlations between flows. Nmap in particular evades detection with this map size due to its rapid spraying of new flows, which quickly evicts older entries before they can be examined. However, the memory requirements of a \textit{flowStats} map with 10240 entries is only 240kB, demonstrating a very low tradeoff between detection accuracy and memory requirements.

\subsection{Testbed Resource Usage}
\begin{figure}
    \centering
    \includegraphics[width=0.9\linewidth]{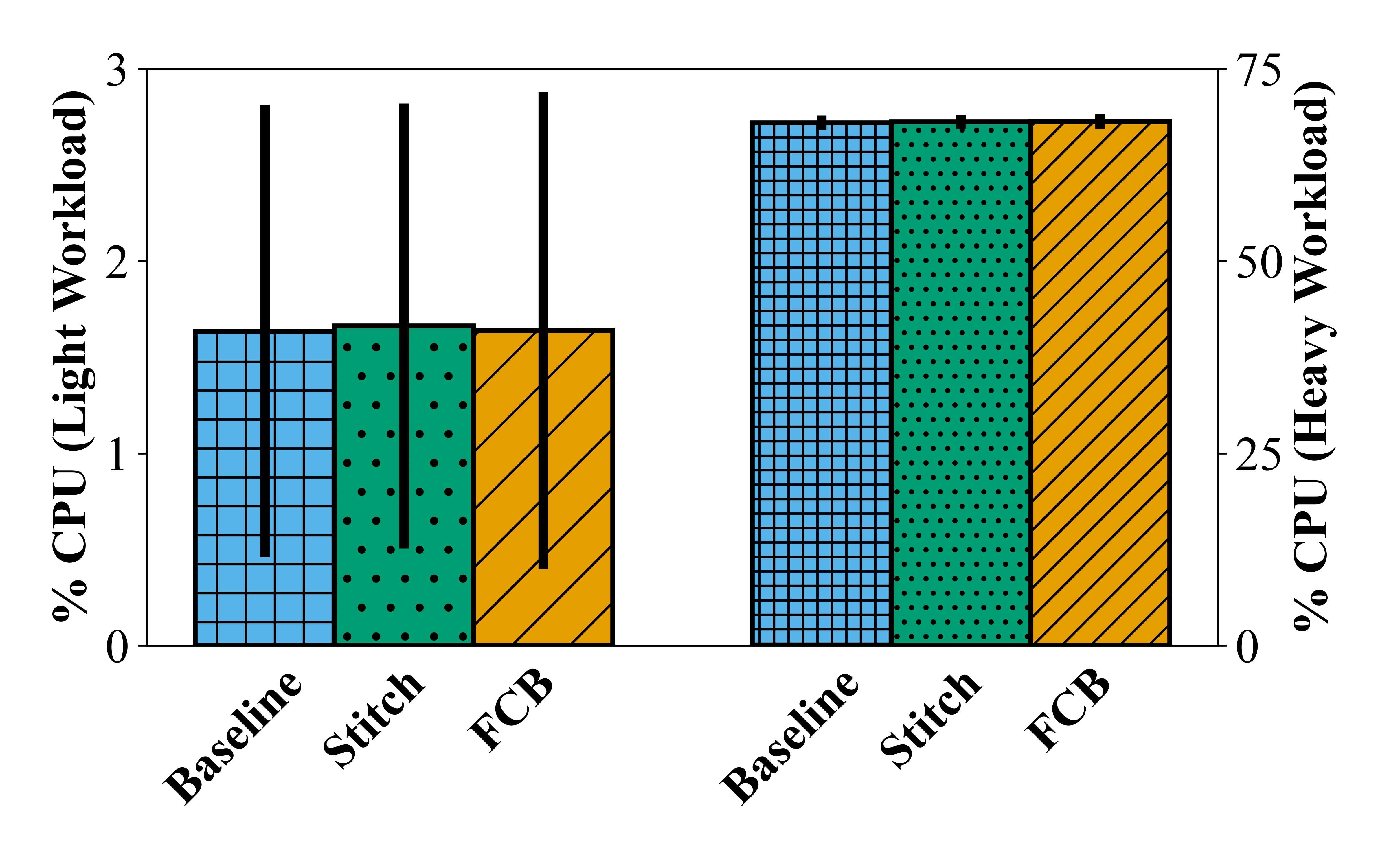}
    \caption{CPU baseline workload compared to the overhead introduced by \sysname{} and the FCB solution under a light workload (left) and a heavy one (right).}
    \label{fig:cpu}
\end{figure}

\textbf{CPU Usage.}
We evaluate the CPU load of \sysname{} in our testbed web server under light and heavy workloads and compare it to the baseline CPU usage with no pivot-detection in place, and with that of the FCB solution. For the light workload, we introduce synthetic user traffic to the web server at a rate of roughly 6 requests per second. For the heavy workload, we use \textit{httperf} to stress all CPUs and generate outgoing traffic at a rate of 16384 requests per second. We take one sample per second for 1500 total samples of CPU load during the experiments. Fig.~\ref{fig:cpu} shows the baseline CPU usage of 1.637\% under a light workload and 67.99\% under a heavy one. With only an additional 0.026\% load with the addition of \sysname{} in a light workload, and 0.097\% in a heavy one, there is no significant addition of CPU overhead from \sysname{} in either scenario. The FCB solution has a similarly light impact with an additional 0.002\% and 0.302\% for light and heavy workloads, respectively. 

\textbf{Memory Usage.}
We analyze the worst-case scenario of memory usage in \sysname{} and compare it with that of the FCB solution. Table~\ref{tab:worst_case} shows the maximum memory usage possible per-component and overall for each system in a 32-bit and a 64-bit environment. We overestimate by assuming a scenario in which the kernel is running the maximum number of processes possible. Thus, \textit{tgids} must be equal to the kernel's pid\_max value to concurrently track all processes. Since pid\_max changes significantly between 32- and 64-bit systems, we provide an analysis for both.
 
\rowcolors{2}{gray!25}{white}
\begin{table}
\centering
\resizebox{\linewidth}{!}{
    \begin{tabular}{|l|c|c|c|c|c|c|c|c|c|}
    \hline
    \rowcolor{gray!50}
    \multicolumn{1}{|c|}{\textbf{System}}   &  \textbf{Network Monitors}  & \textbf{Process Monitor}     & \textbf{Total} \\ \hline
    \sysname{} (32-Bit)                     &  12 MB                    &  2.5 MB                      & 14.5 MB        \\ \hline
    \textit{FCB} (32-Bit)              &  21 MB                      &  0 MB                        & 21 MB          \\ \hline
    \sysname{} (64-Bit)                     &  12.3 MB                    &  320 MB                      & 332.3 MB       \\ \hline
    \textit{FCB} (64-Bit)              &  21 MB                      &  0 MB                        & 21 MB          \\ \hline
    \end{tabular}
}
\caption{Comparison of storage requirements between \sysname{} and the FCB solution. The latter has no process-tracing components while \sysname{} requires space to monitor system processes, the maximum number of which varies across 32- and 64-bit systems.}
\label{tab:worst_case}
\end{table}

The Process Monitor in \sysname{} uses the \textit{tgids} eBPF map to track labeled processes. In a 32-bit system with the upper bound of pid\_max being $\mathrm{2}^{15}$ (32768), the footprint of \textit{tgids} would be a maximum of 2.5MB. This is negligible for modern machines \cite{mem_price}. In a 64-bit Linux system, the upper bound of pid\_max grows to $\mathrm{2}^{22}$ (4194304), and as such the maximum possible size of \textit{tgids} becomes 320MB.

\sysname{}'s Network Ingress and Egress Monitors use \textit{socketIn}, \textit{socketOut}, \textit{flowStats}, \textit{endpoints}, and \textit{output} for tracking information about incoming and outgoing flows. In this case, we imagine the kernel has accepted flows on every possible port (65535 possible TCP ports) and adjust the sizes of \textit{socketIn}, \textit{socketOut}, and \textit{endpoints} to match this number of connections, resulting in map sizes of 5MB each for the socket maps and 786kB for \textit{endpoints}. These values do not change between 32- and 64-bit systems.

The \textit{flowStats} map is more flexible in size and could be customized based on expected traffic levels if memory is a first-class requirement. Allowing for over 43k flow records per 1MB of storage in a 64-bit system, we align the size of \textit{flowStats} with that of \textit{socketIn} and \textit{socketOut} for this analysis, with 65535 entries resulting in 1.3MB and 1.6MB for 32- and 64-bit systems, respectively. Even though the number of records does not change, this size difference occurs in \textit{flowStats} due to different struct padding requirements between systems. The final map, \textit{output}, is a ring buffer used by \sysname{} to communicate between user and kernel space. At 16kB, its footprint is negligible in both 32-bit and 64-bit systems. With a total combined footprint of 11.3MB and 11.6MB for 32- and 64-bit systems, respectively, the Network Monitoring components of \sysname{} have a consistently low memory overhead.

Although comparable in a 32-bit system, the FCB solution has lower space requirements than \sysname{} in a 64-bit system due to the fact that it does not perform process tracing. \sysname{} requires the use of additional space to be able to monitor process activity, while the FCB solution must only keep track of network flows and their characteristics. While storage requirements are lower, the lack of accuracy in a host-based FCB solution outweighs the benefits of a smaller footprint.

\subsection{Live Deployment Environments} 
\textbf{Setup.} We first deployed \sysname{} for 52 days on a live, multi-purpose server in a large faculty network. This server was shared between students and faculty for various needs, including file storage, general compute purposes, and proxy services. It was accessible to roughly 4000 users and handled an average of around 1500 authentication requests per day. It ran Linux Kernel v6.1. 

We then deployed \sysname{} for 32 days in a large academic network on a live web server and on its connected database server. The web server hosted multiple user-managed, publicly accessible websites with an average of around 83000 web visits and 200 SSH connections per day. The connected database server could only be accessed from internal connections for web-related data requests and maintenance. The kernel on both servers was Linux v5.14.

\textbf{Network Traffic.} During the live deployments, most traffic was generated by real users via HTTPS web connections for purposes including database management, remote file management, server management, and web page access, among others. The remaining traffic was generated by automated security and maintenance systems.

\textbf{Pivot Attacks.} During the live deployment on the multi-purpose server, we conducted 85 SSH tunneling attacks using the server as the pivot node. Our attacks originated within the same network as the pivot node and targeted a tertiary general-purpose server. Due to additional privacy and security concerns, we did not perform attacks against the public-facing web server.

\subsection{Live Deployment Results}
\textbf{Multipurpose Server.} We deployed \sysname{} on this server using the detection parameters identified above: $S_w=6500$B, $T_w=1$s, and an endpoint connection threshold of 10. In addition to \sysname{}, the server was protected by traditional security and management features providing authentication, access management, and other services. These measures fail, however, to identify cases where sensitive information or malicious commands are propagated across the system. \sysname{} provides this additional security layer and cooperates with existing measures without any reconfiguration. We perform 85 SSH tunneling attacks against this server to further verify the detection capabilities of \sysname{}. All SSH tunnels were conducted using the same attack and target nodes, meaning that after 10 alerts \sysname{} emitted no further alerts for the remaining attacks due to the endpoint threshold. Although the subsequent attacks after the first 10 did not receive individual alerts, the 10 emitted alerts still implicate all involved nodes and connections due to their identical information.

\sysname{} produced false-positive alerts between experiment days 24 and 49. During this period, several users were port-forwarding through this server to access a database within the internal network. This activity generated an average of 9 alerts per 24-hour day and an overall FPR of 0.005\%. All alerted traffic was tunneling to and from the same service on the same server, and these alerts could be quickly verified by an administrator due to their uniformity. 

Only 5 other false-positive alerts were generated outside of the database port-forwarding alerts. One was related to an automated authentication request, while the remaining four were generated by HTTPS requests made from the server using established tunnels. These five cases represent rare occurrences of highly symmetric traffic generated by authentication or real-world web requests.

\begin{figure}
    \centering
    \includegraphics[width=0.75\linewidth]{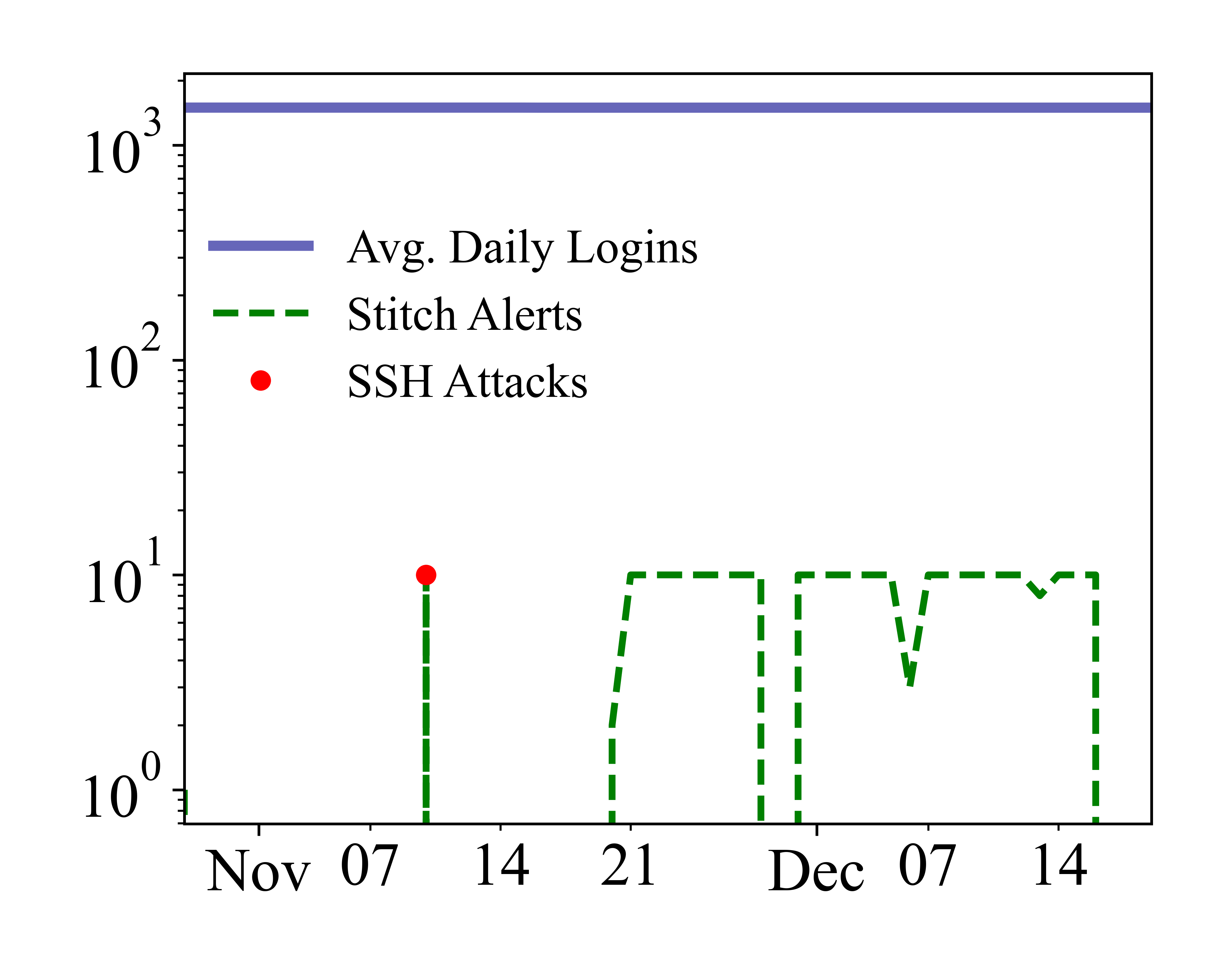}
    \caption{Daily logins, alerts, and attacks that occurred during the live deployment of \sysname{} in a shared faculty server. The later half of the experiment saw increased tunneling activity through a database-access port forwarding service offered by the server.}
    \label{fig:live}
\end{figure}

\begin{figure}
    \centering
    \includegraphics[width=0.75\linewidth]{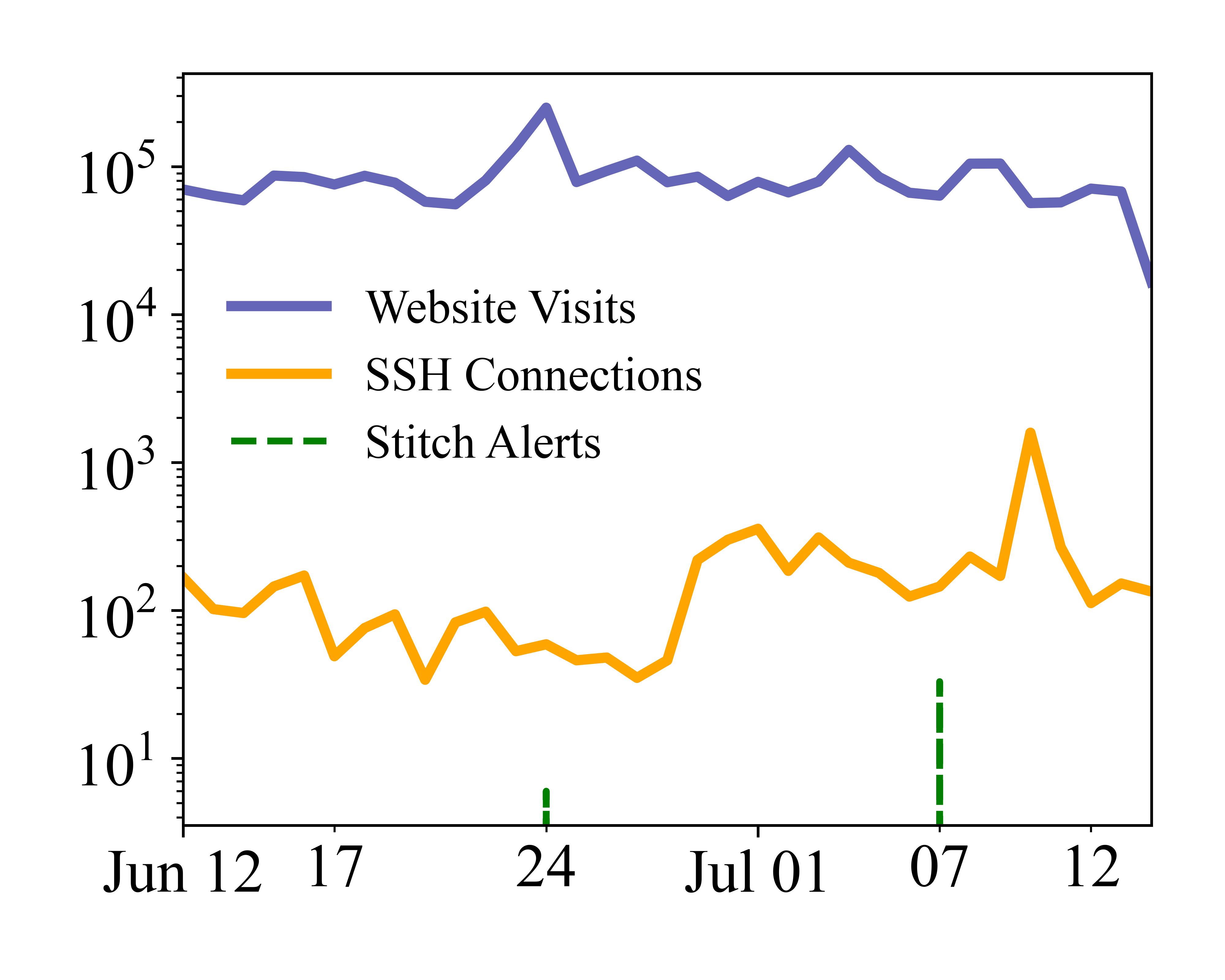}
    \caption{Daily activity and alerts that occurred during the live deployment of \sysname{} in a public-facing web server. \sysname{} flagged tunneling activity on two occasions as a result of database and automated authentication activity.}
    \label{fig:live_2}
\end{figure}

\textbf{Web Server.} Our second live experiment involved deploying \sysname{} both on a public-facing web server and on its directly connected database server for 32 days.  Due to additional security and privacy considerations we did not perform any attacks against these servers. \sysname{} did not generate any alerts on the database server. Due to its strict network access rules, it is unlikely that any receiving process was able to instigate an outgoing flow to a destination other than the connected web server. On the web server, a total of 38 alerts were generated by \sysname{} over the 32-day deployment, shown in Fig.~\ref{fig:live_2} and equivalent to an overall 0.006\% FPR. Of these alerts, 6 were related to accessing services on the connected database server (June 24), and the remaining 32 were related to automated authentication services (July 7). These alerts are notably similar to those generated during deployment on the multi-purpose server, demonstrating consistency in the types of communications or network design decisions that may result in false-positive alerts from \sysname{}. Organization-level network decisions may generate similar consistency in other networks, and allow their respective administrators to adjust \sysname{}'s parameters for optimal functionality in their given context.

Through extensive experiments and live deployments, we demonstrate that the tiered approach of IFC, subsequent characteristics analysis, and endpoint frequency analysis is an effective approach to pivot detection in dynamic environments at scale. While this approach can raise false-positive alerts in some cases (e.g., the database use-case), these occur in small quantity and remain well within network administrators' verification capacities. Moreover, \sysname{} incurs minimal resource overhead, supporting its practicality for real-world deployment scenarios.

\section{Discussion} \label{sec:discussion}

\textbf{Host Security.}
In the event of a complete host compromise, attackers could disable or modify \sysname{} (e.g., remove labels, alter flow records, or suppress alerts) to render it ineffective. However, such modifications require increased effort and create a larger attack footprint, increasing the chances of detection by other security mechanisms (such as anomaly-based IDSes \cite{holmes, prographer}). Mechanisms also exist to secure eBPF programs and maps against malicious root-users \cite{kill_prevention, map_sec}, and adopting them would require only minor changes to \sysname{} and would limit privileged users' ability to terminate the detection agent. More broadly, while eBPF is a safe method of extending OS capabilities, it has known vulnerabilities affecting certain kernel versions (e.g., CVE-2023-2163 \cite{cve_2023}, CVE-2021-41864 \cite{cve_2021}) that could enable malicious kernel code execution or arbitrary memory reads/writes. However, all such vulnerabilities are patched as of this writing, and the efforts to harden eBPF are ongoing and orthogonal to this work.

\textbf{Evasive Attacker.}
Attackers aware of \sysname{}'s detection strategy could pad packets, insert delays into pivot tunnels, or flood new connections to evade detection. However, as with disabling the system itself, these strategies require additional effort and expand the attack footprint, making attacker activity easier to identify for complementary security mechanisms. \sysname{} could be extended with adaptive size and time windows and endpoint counters, or complementary systems could be configured to flag these specific evasive tactics. For example, flagging padded packets entering through the perimeter firewall. \sysname{} should be deployed as part of a defense-in-depth strategy, providing real-time, flexible pivot defense alongside other best-practice tools such as firewalls and auditing systems.

\textbf{Future Work.}
First, a method of providing dynamic map allocation could improve memory usage for any nodes that see frequent downtime but may still need to handle bursts of rapid traffic. Also, to further clarify the legitimacy of pivoting traffic, a packet inspection or additional anomaly identification module (e.g., based on user credentials) could be incorporated into the existing system. Finally, to further respond to dynamic network and system conditions or evasive attackers, \sysname{} could utilize dynamic size and time windows or dynamic endpoint thresholds based on administrator-guided responses to traffic volume. The modular design of \sysname{} allows easy extensions of the detection pipeline, and eBPF is flexible enough to allow for extensions while maintaining the system's high accuracy and lightweight profile. 

\section{Related Work}
\label{sec:related}
\textbf{eBPF in Network Security.} 
Tetragon \cite{tetragon}, Falco \cite{falco}, Tracee \cite{tracee}, and KubeArmor \cite{karmor} are open-source eBPF-based security and observability tools. They use eBPF to monitor system events and raise alerts based on user-defined rules with a focus on container and cloud environments. Although these tools can offer in-depth information about kernel events, they are not designed to correlate network events or draw the inferences from collected data necessary to detect pivoting. Pivot detection requires stateful monitoring of incoming and outgoing network flows and correlation of data across different event types (e.g., network events and syscalls). These tools are primarily to log system events, with a minimal reaction set in the case of Tetragon and KubeArmor.

Works such as \textit{eBPF-LAIN} \cite{ebpf_lain}, \textit{nooBpf} \cite{ebpf_noob}, and the transformer-based approach in \cite{ebpf_transformer} also use eBPF to perform network traffic analysis and security tasks. However, none of these are focused on pivot detection. \textit{eBPF-LAIN} focuses on detecting port-scanning attempts by monitoring only incoming traffic, and so lacks the context required to infer pivoting. \textit{nooBpf} is used to collect network telemetry data, but lacks the internal system context needed to correlate incoming and outgoing traffic. Finally, the system in \cite{ebpf_transformer} uses eBPF to perform feature extraction in a DDoS attack detection system. Attack detection is performed in userspace and requires a trained model, introducing training delays and computation overhead that are not present in \sysname{}. 

\textbf{Pivot Detection.}
Some existing approaches perform pivot detection using only size and time features. \spotlight{} \cite{spotlight} is a distributed in-network pivot-detection solution that provides network-wide coverage using programmable switches to monitor and characterize pivoting traffic in real time, and APIVADS \cite{apivads} is a host-based pivot detection system that similarly characterizes flows based on their sizes and times, but also considers per-packet traffic to identify a pivoting pattern within periodic time windows. We demonstrate that when relying on size and time alone, there is a significant tradeoff between detection accuracy and false positives. By explicitly tracing the flow of information through host machines, \sysname{} achieves a reduced FPR and improved detection accuracy.

\pctl{} \cite{p4control} and \emph{PivotWall} \cite{pivotwall} are network-wide DIFC systems. They use a combination of programmable switches and end-host agents to propagate data labels and perform flow control throughout a network. They prevent pivoting by blocking certain information paths. Although \sysname{} uses process labels to perform internal IFC in a similar manner, the systems differ significantly. DIFC systems require the coordination and participation of all nodes (including hosts, servers, switches, etc.) in a network, while \sysname{} is fully independent and can detect pivoting on a node without the need to track or store external information.

Several works have taken a graph-based approach to pivot detection. Apruzzese et al. \cite{apruzzese} decomposed pivot detection into a temporal graph problem, applying a detection algorithm to network-wide flows within a time window to correlate and rank them using a threat scoring system. \textit{Hopper} \cite{hopper} analyzed network logs to build a graph of common login paths and flag anomalous traversals, while \textit{Holmes} \cite{holmes} and \textit{PROGRAPHER} \cite{prographer} built high-level APT attack summaries from system- and network-level audit logs. Because these approaches rely on offline log analysis and graph construction, they cannot provide real-time responses and their resource requirements can grow rapidly as logs and audit data accumulate.

\section{Conclusion}
\label{sec:conclusion}

This paper presented \sysname{}, a lightweight, real-time, standalone pivot-detection system that operates without any cooperation from surrounding network devices. Using process tracing, flow characteristics, and end-point frequency analysis, \sysname{} employs a tiered pivot detection approach that remains effective even under highly symmetric traffic conditions. Across extensive lab experiments and two live deployments in large academic networks, \sysname{} consistently detected pivoting attacks with an extremely low false-positive rate (0.006\%). In controlled settings, it improved detection effectiveness over existing solutions by +31.49\% while cutting FPR to an average of 0.18\%, demonstrating both robustness and practical accuracy.

\section*{Ethics Considerations}
With the exception of our live deployments, tests were performed with synthetic data in an isolated testbed, mitigating the potential to affect any outside individuals, organizations, or other stakeholders. 

\textbf{Live Deployment.} We identify the following stakeholders that could be potentially impacted by our live experiments on the public-internal servers: server users, IT administrators, and the overarching organization. Live deployments that involve monitoring network traffic and system activity carry the potential to violate privacy principles or service agreements, affecting all identified stakeholders. In cooperation with our organization's IT team, we mitigate these potential harms by:
\begin{itemize}
  
\item Ensuring that the storage and compute resources required by our experiment are negligible and do not impede normal functioning of the server or network.

\item Ensuring that both our attack and monitoring components have no access to private user or organization-owned data. 
\end{itemize}

\noindent
With potential harms sufficiently mitigated, our work contributes positively to APT research and detection techniques without any negative ethical impacts.

\bibliographystyle{IEEEtran}
\bibliography{reference}

\end{document}